\documentclass[prd,twocolumn,preprintnumbers]{revtex4-1}

\usepackage{amsmath}
\usepackage{epsfig}
\usepackage{graphicx}
\usepackage{color}
\usepackage[normalem]{ulem}
 \usepackage{url}
\usepackage[breaklinks, plainpages=false, colorlinks=true, anchorcolor=cyan, linkcolor=red, citecolor=cyan, urlcolor=magenta, bookmarks=false]{hyperref}

\usepackage[caption=false]{subfig}

\begin{document}
\renewcommand{\thefigure}{\arabic{figure}}
\setcounter{figure}{0}

 \def\I{{\rm i}}
 \def\E{{\rm e}}
 \def\D{{\rm d}}

\bibliographystyle{apsrev}

\title{Time-Frequency Analysis of Gravitational Wave Data}

\author{Neil J. Cornish}
\affiliation{eXtreme Gravity Institute, Department of Physics, Montana State University, Bozeman, Montana 59717, USA}

\begin{abstract} 
Data from gravitational wave detectors are recorded as time series that include contributions from myriad noise sources in addition to any gravitational wave signals. When regularly sampled data are available, such as for ground based and future space based interferometers, analyses are typically performed in the frequency domain, where stationary (time invariant) noise processes can be modeled very efficiently. In reality, detector noise is not stationary due to a combination of short duration noise transients and longer duration drifts in the power spectrum. This non-stationarity produces correlations across samples at different frequencies, obviating the main advantage of a frequency domain analysis. Here an alternative time-frequency approach to gravitational wave data analysis is proposed that uses discrete, orthogonal wavelet wavepackets. The time domain data is mapped onto a uniform grid of time-frequency pixels. For locally stationary noise - that is, noise with an adiabatically varying spectrum - the time-frequency pixels are uncorrelated, which greatly simplifies the calculation of quantities such as the likelihood. Moreover, the gravitational wave signals from binary systems can be compactly represented as a collection of lines in time-frequency space, resulting in a computational cost for computing waveforms and likelihoods that scales as the square root of the number of time samples, as opposed to the linear scaling for time or frequency based analyses. Key to this approach is having fast methods for computing binary signals directly in the wavelet domain. Multiple fast transform methods are developed in detail.
\end{abstract}

\maketitle

\section{Introduction}

Gravitational wave data analysis employs many of the standard tools of time-series analysis. With some exceptions, the majority of gravitational wave data analysis is performed in the frequency domain. Notable exceptions include the time domain analysis of pulsar timing array data~\cite{vanHaasteren:2008yh},  time domain low-latency LIGO/Virgo searches~\cite{Messick_2017,Sachdev:2019vvd}, analysis of black hole ringdowns~\cite{Carullo:2019flw,Isi:2019aib}, and wavelet domain searches for gravitational wave transients~\cite{Klimenko:2005xv,Klimenko:2015ypf}. The primacy of frequency domain analyses is due to the advantages it confers for modeling detector noise, under the assumption that the noise properties are at least approximately stationary~\cite{LIGOScientific:2019hgc}. Since most analyses are performed in the frequency domain, significant effort has gone into developing frequency domain waveform models~\cite{Buonanno:2009zt,PhysRevD.77.104017,PhysRevLett.106.241101,Santamaria:2010yb,Hannam:2013oca,Khan:2015jqa,Garcia-Quiros:2020qpx,Chatziioannou:2017tdw,Moore:2018kvz}. 

As ground based interferometers continue to improve in sensitivity, especially at low frequencies, and with the advent of future space based interferometers, the duration that gravitational wave signals spend in the sensitive band of the detectors will increase. The assumption that the noise can be treated as approximately stationary will break down for these long duration signals. It is time for gravitational wave data analysis to leave the frequency domain.

The wavelet domain provides a time-frequency representation of that data that is well suited for modeling non-stationary noise processes. Indeed, for certain discrete wavelet bases, and for noise with statistical properties that vary slowly in time, the noise correlation matrices are diagonal in the wavelet domain, with diagonal entries given by the dynamic, or evolutionary power spectrum~\cite{doi:10.1111/j.2517-6161.1965.tb01488.x} of the noise, $S(f,t)$.  The Wilson-Daubechies-Meyer (WDM) wavelet basis~\cite{Necula_2012} employed by the coherent WaveBurst algorithm~\cite{Klimenko:2005xv,Klimenko:2015ypf} is a good choice for more general gravitational wave analyses. The WDM wavelet basis provides excellent frequency separation and a uniform tiling in time and frequency. The coherent WaveBurst algorithm uses the WDM basis to represent both data and signals -- all data conditioning, power spectral estimation and likelihood calculations are carried out in the wavelet domain.  In contrast to the cWB algorithm, which reconstructs signals as a collection of WDM wavelets, here the WDM basis simply replaces the Fourier basis - any signal model, including waveform templates derived from general relativity, can be mapped to the WDM basis. A key ingredient for the wavelet domain analysis is having waveform models that can be computed directly and efficiently in the wavelet domain. For the most commonly encountered type of gravitational wave signal - produced by the merger of compact binary stars on quasi-circular orbits - it turns out that the waveforms can be computed much more efficiently in the wavelet domain than in the time or frequency domains. The computational cost in the wavelet domain scales as the square root of the observation time, as opposed to linearly with time as is the case in the time and frequency domains. Methods are introduced for computing binary inspiral-merger-ringdown (IMR) waveforms directly and efficiently in the wavelet domain. Codes for computing the WDM transform, with examples of the fast signal transform methods can be found at \url{https://github.com/eXtremeGravityInstitute/WDM_Transform}.
%{\tt https:\slash\slash github.com\slash eXtremeGravityInstitute\slash WDM\_Transform}.

\section{Motivation}

Gravitational wave data is recorded as a collection of time series from a network of detectors. Evenly spaced data from interferometric detectors are amenable to the standard techniques of time series analysis, such as linear filtering and spectral analysis~\cite{Finn_1992,Allen_2012}. A key quantity that appears in both Frequentist and Bayesian approaches to time series analysis is the noise weighted inner product, $(a|b)$, of time series $a$ and $b$, given by the expression
\begin{equation}\label{nwip}
(a|b) = {\bf a}^\dag \cdot {\bf C}^{-1} \cdot {\bf b} = a_i \, C_{ij}^{-1} \, b ^j \, ,
\end{equation}
where $C_{ij}$ is the noise correlation matrix. Here the sums run over the individual data samples from each detector in the array. Important examples where this inner product is used include the match-filter signal-to-noise ratio
\begin{equation}
\rho(h) = \frac{ (d|h) }{\sqrt{(h|h)}} \, ,
\end{equation}
where $d$ is the data and $h$ is a waveform model for the signal, and the likelihood function for Gaussian noise
\begin{equation}
p(d|h) = \frac{1}{\sqrt{ {\rm det}(2 \pi {\bf C})}} e^{-\frac{1}{2} ({\bf d} - {\bf h})^\dag \cdot {\bf C}^{-1} \cdot ({\bf d} - {\bf h})} \, .
\end{equation}

Several challenges present themselves when computing these quantities. The first is that the noise correlation matrix is not know in advance and has to be estimated from the data. The second is the computation cost of evaluating the inverse of the noise correlation matrix and the cost of evaluating the sum that appears in (\ref{nwip}). Interferometer noise is usually highly colored, with both red and blue components, leading to strong correlations between data samples in the time domain. With $N$ data samples, the cost of inverting the noise correlation matrix in the time domain scales as $N^3$, while the cost of the sums scale as $N^2$. It is also very difficult to estimate the $N^2$ elements of the correlation matrix from the $N$ observed data points. One solution to these problems is to apply a linear transform to the time series: $\tilde{{\bf a}} = {\bf Q} \cdot {\bf a}$ that diagonalizes the noise correlation matrix:
\begin{equation}\label{nwip2}
(a|b) = \tilde{\bf a}^\dag \cdot \left( {\bf Q}^\dag {\bf C}^{-1} {\bf Q} \right) \cdot \tilde{{\bf b}} = \frac{ {\tilde a}_i  {\tilde b}_i}{\sigma^2_i}
\end{equation}
where $\sigma^2_i$ are the diagonal entries of the transformed noise correlation matrix. The computation cost of the matrix inversion and sum now scales linearly with the number of data points $N$. The cost of the matrix inversion has been traded for the cost of finding the linear transform ${\bf Q}$. If the data happen to be stationary, that is, if the correlations between the time domain samples only depends on the time interval between samples, and not the sample times themselves, then ${\bf Q}$ can be found using several different methods. One method is to fit an auto regressive (AR) or auto regressive moving average (ARMA) model to the data, then use the inverse of this model to pre-whiten the data~\cite{Cuoco:2001wz,Cuoco:2000gv}. A simpler method is to Fourier transform the data, since for stationary time series the noise correlation matrix is diagonal in the frequency domain, with $\tilde{C}_{ij} = S(f_i) \delta_{ij}$, where $S(f)$ is the Fourier power spectrum. The power spectrum can be estimated by a variety of methods, including classical methods such as Welch averaging, or Bayesian methods that use parametric~\cite{PhysRevD.91.084034} or semi-parametric~\cite{PhysRevD.92.064011} modeling.

In reality, the noise encountered in gravitational wave data analysis is non-stationary~\cite{Aasi:2012wd,Aasi:2014mqd,TheLIGOScientific:2016zmo}, which significantly complicates the estimation of the the noise correlation matrix and the calculation of noise weighted inner products. If the noise properties vary slowly in time, and are approximately constant over the time a signal is in the sensitive band of the detectors, then it is possible to treat the data as stationary in the short data segment containing the signal. The data is then analyzed in short chunks, with the power spectral density (PSD) re-computed for each chunk. This approach becomes increasingly inaccurate as the duration of the signals increases. For example, binary neutron star signals will be in-band for minutes or hours as the low frequency sensitivity of ground based detectors improves. For space based detectors such as LISA, the signals will be in band for months or years. Even if the LISA instrument noise is perfectly stationary (unlikely!), the residual signal from unresolved galactic binaries will be highly non-stationary~\cite{Adams:2013qma}.

The usual frequency domain approach fails for non-stationary data as the noise correlation matrix is no longer diagonal. When the variations in the noise properties are small, the frequency domain noise correlation matrix becomes band-diagonal. A $N\times N$ band diagonal matrix with bandwidth $n$ can be decomposed into a product of upper and lower triangular matrices in order $n^2 N$ operations~\cite{KILIC2013126}. These triangular matrices can be inverted in order $N^2$ operations, resulting in the total cost of the inverse scaling as $N^2$.  Since the inverse of a band diagonal matrix is generally not band diagonal, the cost of the sum and the inverse in (\ref{nwip}) will both scale as $N^2$. Added to this increased computational cost is the challenge of estimating the ${\cal O}(n N)$ off-diagonal elements of the noise correlation matrix. 

Non-stationary noise has been shown to impact frequency domain matched filter searches for gravitational waves, resulting in significant losses of efficiency for long duration signals~\cite{Zackay:2019kkv,Mozzon_2020}. It was found the loss of efficiency could be mitigated to some extent by applying PSD drift corrections~\cite{Zackay:2019kkv} or by re-ranking the search statistic~\cite{Mozzon_2020}. These work-arounds avoid the more costly step of using a non-diagonal noise correlation matrix, but it is unclear if these approaches could be adapted for use in other settings, such as Bayesian parameter estimation.

An alternative approach that has been proposed is to generalize the ARMA pre-whitening to allow for non-stationary noise~\cite{Cuoco:2004tr,DAHLHAUS2012351}. It is unclear how the computational cost of such an approach will scale given the added complexity of the non-stationarity, but it is likely to be significantly more expensive than the standard frequency space analysis. As a point of comparison, fitting an $AR(p)$ model to a stationary time requites order $pN$ operations to compute the $p$ lagged auto-corrrelation functions, and order $p^2$ operations to solve the Yule-Walker equations to find the model coefficients. For typical LIGO-Virgo spectra the model order $p$ will be order $N$. Note that both the data and the waveform templates have to be filtered by the model at a cost that scales as $pN$. The computational cost for non-stationary time series will likely be significantly higher.

Given that the noise properties vary with time, it is natural to consider wavelet based time-frequency methods. A discrete wavelet (or wavelet wavepacket) transformation provides a linear mapping from the time domain to the two-dimensional time-frequency domain indexed by time $n$ and frequency $m$. With an appropriate wavelet basis, and for a wide array of locally non-stationary processes~\cite{dahlhaus1997,doi:10.1098/rsta.1999.0445}, the wavelet noise correlation matrix is diagonal:
\begin{equation}\label{lswp}
C_{(nm)(n'm')} = \delta_{n n'} \delta_{m m'} S(t_n,f_m) \, ,
\end{equation}
where $S(t,f)$ is the dynamic, or evolutionary, power spectrum~\cite{doi:10.1111/j.2517-6161.1965.tb01488.x}. Given the diversity of non-stationary processes, there is no general proof that the wavelet correlation matrix will always be diagonal, but proofs exists for certain cases. To understand, at least heuristically, why the wavelet noise correlation matrix might be diagonal, it is helpful to recall why the Fourier domain noise correlation matrix is non-diagonal for non-stationary processes. In the Fourier domain, information about when an event occurred is encoded in the Fourier phase, not the Fourier spectrum $S(f)$. For stationary processes, the phase in each Fourier bin is uncorrelated, and the noise correlation matrix is diagonal. For non-stationary processes the time dependence leads to correlations in the Fourier phases, and this gets encoded in the off-diagonal elements of the noise correlation matrix. In contrast, the wavelet domain has an evolutionary spectrum $S(t,f)$ that can account for certain types of non-stationarity, so the time dependence does not have to be encoded in the off-diagonal elements of the noise correlation matrix. The noise correlation matrix has been shown to be orthogonal for wavelet-like bases, such as certain types of windowed Fourier transforms~\cite{mallat1998,10.2307/2670296}. The WDM wavelet family~\cite{Necula_2012} used in the analyses in this paper can be computed using an FFT with a specially chosen window function. Additionally, discrete wavelet bases that are well localized in frequency, such as the ones used here, have been shown to yield approximately diagonal correlation matrices for locally stationary red noise processes~\cite{MaciasPerez:2005uz}. Turning things around, the condition (\ref{lswp}) has been used to define a class of locally stationary wavelet processes (LSWP)~\cite{doi:10.1111/1467-9868.00231,doi:10.1111/jtsa.12230} that can be used to simulate and model non-stationary time series. 

\section{Discrete Wavelet Wavepackets}

Time-frequency representations of gravitational wave data using oversampled continuous wavelets (e.g. Q-scans~\cite{Chatterji:2004qg})  provide nice visualizations of the gravitational wave signals, and are also widely used to identify noise transients. The oversampling makes for nice smooth images, but the resulting correlations between pixels excludes these representations from being used in quantitative analyses. In contrast, orthonormal discrete wavelet transform produce poor visualizations, but provide a critically sampled representation of the data with a lossless inverse and appealing statistical properties for handling colored, non-stationary noise.

While there are many choices of discrete wavelet bases, there is one that is ideally suited for gravitational wave data analysis, the Wilson-Daubechies-Meyer (WDM) wavelet wavepackets introduced by Necula {\it et al}~\cite{Necula_2012}. The WDM representation produces a uniform time-frequency grid, as opposed to the more common Dyadic representations which shrink in time as they go up in frequency. The WDM wavelets provide excellent spectral separation, with a band-pass that maintains the same shape across all frequencies. Unlike typical wavelet transform that are computed using repeated application of filer function, the WDM transform can be computed using windowed fast Fourier transforms. Indeed, the WDM transform can be thought of as a short-time Fourier transform (SFT)with a well designed window function. Because of the close relationship to SFTs, the WDM transform can be computed efficiently using FFTs, and the time translations can be incorporated using Fourier domain phase shifts~\cite{Necula_2012}.

\begin{figure}[htp]
\includegraphics[width=0.5\textwidth]{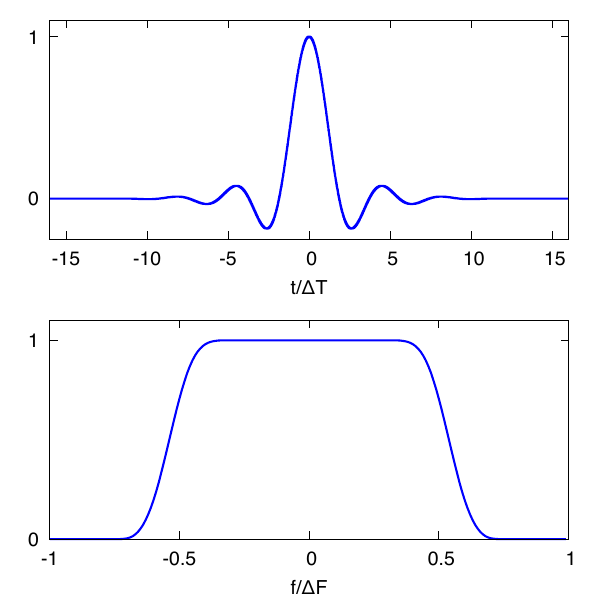} 
\caption{\label{fig:window} The window function for the WDM wavelets for the choice $d=4$, $A = \Delta \Omega/4$ and $B = \Delta \Omega/2$. For this choice of window parameters the wavelets are better localized in frequency than they are in time. Here the overall normalization is arbitrary. }
\end{figure}

The WDM wavelet wavepackets form a complete, orthogonal basis that can be used to faithfully represent any time series:
\begin{equation}
x[k] = \sum_{n = 0}^{N_t-1}\sum_{m=0}^{N_f-1}  w_{nm} \, g_{nm}[k] \, .
\end{equation}
A time series of duration $T$ sampled at a cadence $\Delta t = T/N$ can be represented by a rectangular grid with $N_t$ time slices of width $\Delta T$ and $N_f$ frequency slices of width $\Delta F$:
\begin{eqnarray}
\Delta T &= &N_f \, \Delta t \nonumber \\
\Delta F &= &\frac{1} { 2 \Delta t\, N_f} \, .
\end{eqnarray}
Each of the $N = N_t N_f$ cells has area $\Delta T \Delta F = 1/2$. The time and frequency resolution can be varied by changing $N_f$, such that as $N_f \rightarrow N/2$ the expansion approaches a Fourier series, and as $N_f \rightarrow 1$ the expansion approaches the original time series.  The orthogonality condition
\begin{equation}
 \sum_{k = 0}^{N-1}g_{nm}[k] g_{pq}[k] = \delta_{np} \delta_{mq}
\end{equation}
can be used to find an expression for the expansion coefficients:
\begin{equation}\label{coeff}
w_{nm} =  \sum_{k = 1}^{N} x[k] g_{nm}[k] \, .
\end{equation}
The WDM wavelets are actually a wavelet family, with their precise shapes controlled by three parameters, the half-width of the flat response in frequency, $A$, the width of the frequency roll-off $B$, and the steepness of the frequency roll-off $d$, subject to the constraint $2A+B = \Delta \Omega$, where $\Delta\Omega = 2\pi \Delta F$. Accordingly the wavelets are defined in the frequency domain:
\begin{eqnarray}\label{wavelets}
&& {\tilde g}_{nm}(\omega) = e^{-i n \omega \Delta T} \left(C_{nm}\Phi(\omega - m \Delta \Omega) \right. \nonumber \\
&&\hspace*{1.1in} \left. + C^*_{nm} \Phi(\omega + m \Delta \Omega)\right)  \, ,
\end{eqnarray}
where $C_{nm} = 1$ for $(n+m)$ even and  $C_{nm} = i$ for $(n+m)$ odd. Note that the $m=0$ frequency band needs to be handled a little differently from the others - see Ref.~\cite{Necula_2012} for details. The Meyer~\cite{meyer1990ondelettes} window function $\Phi(\omega)$ is defined:
\begin{equation}
\Phi(\omega) =  \left\{\begin{array}{lr}
       \frac{1}{\sqrt{\Delta \Omega}} , &  |\omega| <  A\\
         \frac{1}{\sqrt{\Delta \Omega}}\cos\left[ \nu_d\left( \frac{\pi}{2} \frac{|\omega| -A}{B}\right) \right], & A\leq |\omega| \leq A+B
        \end{array}\right. \, ,
\end{equation}
where $\nu_d(x)$ is the normalized incomplete Beta function
\begin{equation}
\nu_d(x) = \frac{ \int_0^x y^{d-1} (1-y)^{d-1} dy}{\int_0^1 y^{d-1} (1-y)^{d-1} dy} \, .
\end{equation}
Figure~\ref{fig:window} shows the window function in the time domain, $\phi(t)$, and frequency domain $\Phi(f)$ for $d=4$, $A = \Delta \Omega/4$ and
$B = \Delta \Omega/2$. Formally $\phi(t)$ extends indefinitely in time, but in practice it can be truncated without incurring a significant error. For the example shown in Figure~\ref{fig:window}, the time domain window is truncated beyond $\pm q \Delta T$ with $q=16$. The usual rules for time-frequency localization pertain - a well localized frequency response (large $d$ and small $B$) comes at the cost of a longer window in the time domain. 

Using the definition of the WDM wavelets (\ref{wavelets}) and the expression (\ref{coeff}), the expansion coefficients are given by
\begin{equation}\label{fullt}
w_{nm} =  \sqrt{2} \Delta t \Re\, C_{nm} \sum_{k = -K/2}^{K/2-1}  e^{i\pi k m/N_f} x[n N_f + k]\phi[k].
\end{equation}
Here $K= 2 q N_f$ is the width of the window function, where $q > 1$ is a multiplier that sets the size of the window relative to the temporal extent of the time-frequency pixels. Defining the windowed discrete Fourier transform of $x[n N_f+k]$ as
\begin{equation}\label{fulltx}
X_{n}[j] =  \sum_{k = -K/2}^{K/2-1}  e^{2\pi i k j /K} x[n N_f + k]\phi[k],
\end{equation}
we have
\begin{equation} 
w_{nm} =  \sqrt{2} \Delta t \Re\, C_{nm} X_{n}[mq] \, .
\end{equation}
In other words, the WDM transform of a time series can be computed using $N_t$ windowed short Fourier transforms of length $K$, then downsampling to extract every $q^{\rm th}$ coefficient. The cost of these operations scale as $N_t K \ln(K) =  2q N \ln(2q N_f)$, as compared to the $N \ln(N)$ cost of a standard fast Fourier transform of the data. For the window parameters used in Figure~\ref{fig:window}, $q=16$ and the WDM transform is more than an order of magnitude slower than a straight FFT.

The additional cost of the time-domain WDM transform is due to the length of the time domain window function. Since the window function is more compact in the frequency domain, it turns out to be faster to compute the transform using the discrete Fourier transform of the data $X[l]$:
\begin{equation}\label{fullf}
w_{nm} =  \sqrt{2} (-1)^{nm}  \Re\, C_{nm}  x_{m}[n]\, ,
\end{equation}
where
\begin{equation}\label{fullfx}
x_{m}[n] = \sum_{l = -N_t/2}^{N_t/2-1}  e^{-2\pi i l n/N_t} X[l+mN_t/2]\Phi[l].
\end{equation}
Note that the discrete Fourier samples are evaluated at $f = l \Delta f$, where $\Delta f = 1/T$, and $T = N \Delta t$ is the total span of the data. The $x_{m}[n]$ can be computed using a FFT of length $N_t$ for each $m$. The total computational cost of the frequency domain version of the transform is then $N_f N_t \ln(N_t) = N \ln(N_t)$. This is less than the cost of the FFT of the full time series that has to be done before computing the frequency domain WDM transform. Performed this way, the total cost of the WDM transform is less than twice the cost of the full FFT.

While the gravitational wave data has to be transformed directly using the above methods, gravitational wave signals for which analytic models exist can be evaluated directly in the wavelet domain at a much reduced computational cost using techniques described in later sections.

\subsection{Likelihood, Fisher Matrix and Searches}

The Whittle log likelihood can be evaluated in the wavelet domain:
\begin{equation}\label{whittle}
\ln p(d|h) = -\frac{1}{2} \sum_{nm} \left( \ln (2 \pi S_{nm}) + \frac{(d_{nm}-h_{nm})^2}{S_{nm}} \right) 
\end{equation}
where $S_{nm}$ is the evolutionary power spectrum. For stationary noise, the evolutionary power spectrum can be related to the usual power spectral density $S(f)$ using equations (\ref{fullf}) and (\ref{fullfx}):
\begin{equation}\label{psd}
S_{nm} = \sum_{l = -N_t/2}^{N_t/2-1}  S[l + m N_t/2] \Phi[l]^2 \approx S(f_m) \Delta F
\end{equation}
where $f_m = m N_t/2$ is the central frequency of the wavelet band, and $ \Delta F$ is the width of the band. Note that the $S_{nm}$ are dimensionless.

The Fisher information matrix, defined as $\Gamma_{ij} = -\partial_i \partial_j {\rm E}[\ln p(d|h)\vert_{ML}]$, is given by
\begin{equation}\label{fisher}
\Gamma_{ij} =  \sum_{nm}   \frac{h_{nm,i} h_{nm,j}}{S_{nm}}  \, .
\end{equation}
The wavelet domain Fisher matrix avoids taking derivatives of rapidly oscillating waveforms, and is thus more stable than versions based on direct numerical evaluation in the frequency domain or time domain. 

The signal-to-noise statistic $\rho$ is given by
\begin{equation}\label{rho}
\rho(h) =  \sum_{nm}   \frac{\hat{h}_{nm} d_{nm}}{S_{nm}}  \, ,
\end{equation}
where $\hat{h}_{nm} = h_{nm}/(h|h)^{1/2}$ is a unit normalized waveform template. The waveform templates typically depend on multiple parameters, including intrinsic parameters such as the masses and spins of a binary system, in addition to extrinsic parameters such as the overall amplitude, phase and reference time. The later three parameters are usually analytically maximized over in a search~\cite{LIGOScientific:2019hgc} resulting in considerable computational savings. The maximization over time exploits the fact that time shifts $\delta t$ appear as linear phase shifts $\exp(2\pi i f\delta t)$ in the frequency domain. Similar tricks can be used in the wavelet domain, after taking care of some complications cased by the evolutionary power spectrum $S_{nm}$. When the noise is non-stationary $S_{nm} \neq S_{lm}$ for $n\neq l$, and the time-translation invariance of the template normalization is broken. The time shifted normalization,
\begin{equation}\label{norm}
(h|h)_k=  \sum_{nm}   \frac{h_{nm} h_{nm}}{S_{(n+k)m}}  \, ,
\end{equation}
is typically a slowly varying quantity that can be computed for a small number of time shifts $k$ and interpolated for intermediate values. The computational cost of evaluating $(h|h)_k$ for a binary system is low since the $h_{nm}$ are typically only non-zero for order $\sqrt{N}$ points (more precisely, for a single harmonic the waveforms are non-zero for between $\sim N_t$ and $\sim q N_f$ points). The total cost to compute $(h|h)_k$ is less than $N$ for waveforms with a single dominant harmonic. The inner product $(h|d)$ can be maximized with respect to the overall phase $\phi_0$ using the usual method of quadrature phase waveforms $h(\phi_0=0)$ and $h(\phi_0=\pi/2)$. The maximization over the time shift $\delta t$ can be performed in a number of ways. The simplest and fastest method is to over-whiten the data using the evolutionary spectrum: $\bar{d}_{nm} = d_{nm}/S_{nm}$, then inverse wavelet transform the data to the frequency domain $\bar{d}(f)$. The time maximization can the be done by inverse Fourier transforming the quantity
\begin{equation}\label{zeq}
\tilde z = 4 \tilde h(f) \bar{d}(f) \Theta(f)
\end{equation}
where $\Theta(f)$ is the Heaviside step function. The signal-to-noise statistic is then given by $\rho(\delta t) = |z(\delta t)|/(h|h)^{1/2}_k$ with $k = [\delta t/\Delta T]$. The cost of this procedure is roughly two to three times that of a purely frequency domain implementation due to the extra Fourier transforms involved in computing $d_{nm}$ then $\bar{d}(f)$, and the additional cost of computing the time-shift template normalization $(h|h)_k$. On the other hand, if the data is non-stationary on the time scale of the signal, the wavelet based search statistic will be more robust. This method shares some similarities to the PSD drift correction approach~\cite{Zackay:2019kkv}.

The approach described above can be incorporated in existing search pipelines with minimal changes to the existing codes. The frequency domain waveforms that are currently being used can be transformed to the wavelet domain using equation (\ref{fullf}). While not as fast as the direct wavelet domain methods described in later sections, the cost is only a factor of two larger than the standard frequency domain calculation, and requires no changes to the waveform libraries that have been developed over many years. The main new ingredient is the wavelet domain whitening and associated estimation of the evolutionary spectrum $S_{nm}$. Methods for implementing these steps are described in the next section.

The time shift maximization can also be performed directly in the wavelet domain by treating $h_{nm}$ and  $\bar{d}_{nm}$ as time series in $n$ and applying an FFT to each:
\begin{equation}
H_{m}[k] = \sum_{n=0}^{N_t-1}  e^{2 \pi i n k/N_t} h_{nm},
\end{equation}
and similary for $\bar{d}_{nm}$. The inverse FFT, $z[p]$ of the quantity
\begin{equation}
Z[k] = \sum_{m=0}^{N_f-1} H_{m}[k] \bar{D}^*_{m}[k]
\end{equation}
yields $\rho(p \Delta T)= |z[p]|$. The cost of the $N_f$ FFTs used to compute $H_{km}$ and $\bar{D}_{km}$ scale as $N_t \ln N_t$, so the total cost of computing $Z_k$ scales as $N \ln N_t$. The limitations of this approach are that the time maximization is quantized in units of $\Delta T = N_f \Delta t$, which is much coarser than the $\Delta t$ resolution of the usual time maximization (\ref{zeq}). The time resolution can be improved by generating $p$ reference templates spaced by $\Delta T/p$ and repeating the maximization for each shifted template. The cost of achieving the full $\Delta t$ resolution is then $N_f$ times greater than the usual Fourier domain approach, though using lower time resolutions may be acceptable.  With WDM wavelets the time maximization yields a maximization across frequency bands that may be useful for computing banded time-frequency statistics~\cite{PhysRevD.71.062001}. Using equations (\ref{fullf}) and (\ref{fullfx}), the signal-to-noise statistic becomes
\begin{eqnarray}\label{wdmshift}
\rho(h) &=& \frac{N_t}{2}  \sum_{m=0}^{N_f}   \sum_{l = -N_t/2}^{N_t/2 - 1} \left( H[l+ m N_t/2] \bar{D}^*[l+ m N_t/2] \right. \nonumber \\
&& \quad  \left. + H^*[l+ m N_t/2] \bar{D}[l+ m N_t/2] \right) \Phi[l]^2
\end{eqnarray}
where $H[j]$ denotes the DFT of $h$ and $\bar{D}[j]$ denotes the DFT of $\bar{d}$. The above expression follows from performing the sum over the time index $n$ and using the identity
\begin{equation}
\sum_{n=N_t/2}^{N_t/2-1} e^{-2\pi i (l-k)  n/N_t} = N_t \delta_{lk} \, .
\end{equation}
Introducing the time-shifted template
\begin{equation}
H_{\delta t}[j] = H[j] e^{-2\pi i j \delta t/T},
\end{equation}
and setting $\delta t = p \Delta T$, we have
\begin{equation}\label{rhowdm}
\rho(p \Delta T) = \sum_{m=0}^{N_f} (-1)^{pm} u_{m}[p]
\end{equation}
where $u_{m}[p]$ is given by the inverse FFT of
\begin{equation}
U_{m}[l]= H[l+ m N_t/2] \bar{D}^*[l+ m N_t/2]  \Phi[l]^2 \, .
\end{equation}
The cost of the banded time maximization scales as $N_f N_t \ln N_t$. The time resolution of the maximization can be increased by setting
$\delta t = p \Delta T/a$ for some integer $a$, and extending the sum over $l$ in (\ref{wdmshift}) to run from $-aN_t/2$ to $aN_t/2-1$ and writing
\begin{equation}
\rho(p \Delta T) = \sum_{m=0}^{N_f} e^{-\pi i pm /a}  u_{m}[p]
\end{equation}
Since $\Phi[l]$ is zero for $|l| > N_t/2$ the $U_{m}[l]$ are simply being zero padded to increase the temporal resolution. Note that for $a=N_f$ the temporal resolution is equal to the original sample cadence $\Delta t$. Using a standard FFT, the cost of the banded time maximization scales as $a N \ln (aN_t)$, but it may be possible to achieve faster speeds using sparse FFTs that are optimized for almost empty arrays. 

 \begin{figure}[htp]
\includegraphics[width=0.48\textwidth]{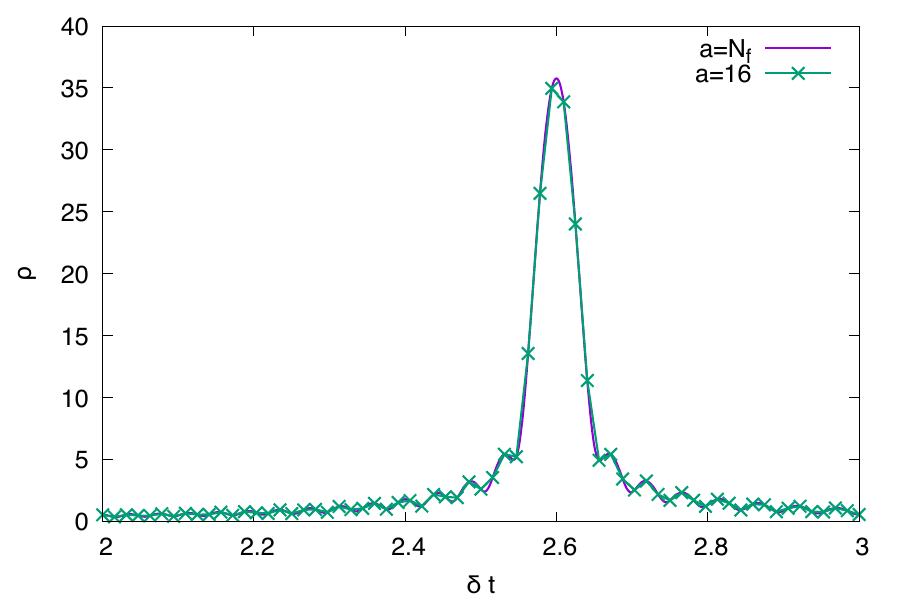} 
\caption{\label{fig:tmax} The signal-to-noise time series $\rho(\delta t)$ for a chirplet waveform with $t_p = 2.6$ s. The curve labeled $a=N_f$ used the standard Fourier domain time maximization with time resolution $\Delta t = 1/1024$ s, while the curve labeled $a=16$ used the WDM banded maximization with resolution $\Delta T/16= 1/60$ s.}
\end{figure}

Figure~\ref{fig:tmax} compares the standard Fourier domain time maximization to the banded WDM time maximization. The simulation used a chirplet waveform (defined in equation (\ref{chirpt}) of section~\ref{wtc}) with central time $t_p=2.6$ s, central frequency $f_p = 416$ Hz, duration $\tau = 3.23$ s, and frequency spread $\gamma = 13$ Hz, sampled at cadence $\Delta t = 1/1024$ s. The WDM transform used $N_f=256$, $N_t=32$ and $\Delta T=0.25$ s.

\section{Glitches, Gaps and other Gremlins}

{\it ``We don't need your fancy statistics, we just need to build better detectors''} -- Rainer Weiss, to the author, at The Tenth Harvard-Smithsonian Conference
on Theoretical Astrophysics, May 2018.\\

The noise in an ideal detector would be stationary and Gaussian, with a level set by irreducible quantum uncertainty. In reality, the noise encountered in contemporary gravitational wave detectors is neither stationary nor Gaussian due to frequent short duration noise transients (glitches), in addition to longer duration drifts in the system~\cite{LIGOScientific:2019hgc}. The data also has many gaps, and the dynamic noise spectra are littered with dozens of sharp spectral lines that wander around in frequency and vary in amplitude. Performing analyses in the wavelet domain helps reduce the impact of many of these issues, but the main benefit is in dealing with relatively slow changes in the noise properties. Methods such as Bayesian data augmentation, edge wavelets, and glitch subtraction are still needed to mitigate the impact of gaps and short duration noise transients. The coherent WaveBurst algorithm~\cite{Klimenko:2005xv,Klimenko:2015ypf} uses the WDM transform for noise modeling, including for power spectrum estimation and spectral line removal~\cite{Tiwari:2015ofa}.

The Heisenberg-Gabor uncertainty relation~\cite{gabor} (which is saturated for discrete wavelets: $\Delta T \Delta F = 1/2$) implies a trade-off between spectral and temporal resolution in the wavelet transform. The time resolution $\Delta T$ should be short compared to the timescale of the non-stationarity of the data, while the frequency resolution $\Delta F$ should be small compared to the scale on which the average power spectrum $S(f)$ varies. If the data are highly non-stationary and also contains sharp spectral features it can be difficult to satisfy both criteria simultaneously with a single choice of resolution, and multi-resolution approaches may be required (e.g. first removing spectral lines using a transform with large $\Delta T$, then taking care of the non-stationarity using a transform with a smaller $\Delta T$).

\subsection{Drifts in the noise spectrum}

Non-stationary noise comes in myriad forms. Stationary noise, heuristically speaking, is generated by a random process with statistical properties that are unchanging in time. Wide sense stationary (WSS) noise (sometimes called weak sense stationary) requires that the first (mean) and second (covariance) moments of the random process are unchanging in time. For WWS processes the autocorrelation between data at times $t_1$ and $t_2$ depends only on the lag $\tau = |t_1-t_2|$. In practice, noise is never exactly WSS. A more realistic approximation is the concept of locally stationarity noise (LS), where the statistical properties of the noise process vary slowly with time~\cite{dahlhaus1997,10.2307/41806549}. The auto-correlation properties of LS processes can be described in terms of dynamic, or evolutionary, spectra $S(f,t)$~\cite{doi:10.1111/j.2517-6161.1965.tb01488.x}. In contrast to stationary processes, where the frequency domain auto-correlation (power spectrum) is non-zero, non-stationary processes produce non-vanishing cross-correlations between frequencies~\cite{mallat1998}. The traditional way to estimate $S(f,t)$ is using periodograms of short time Fourier transforms. Scalograms computed from discrete wavelet transforms are a powerful tool for estimating evolutionary spectra.

\begin{figure}[htp]
\includegraphics[width=0.48\textwidth]{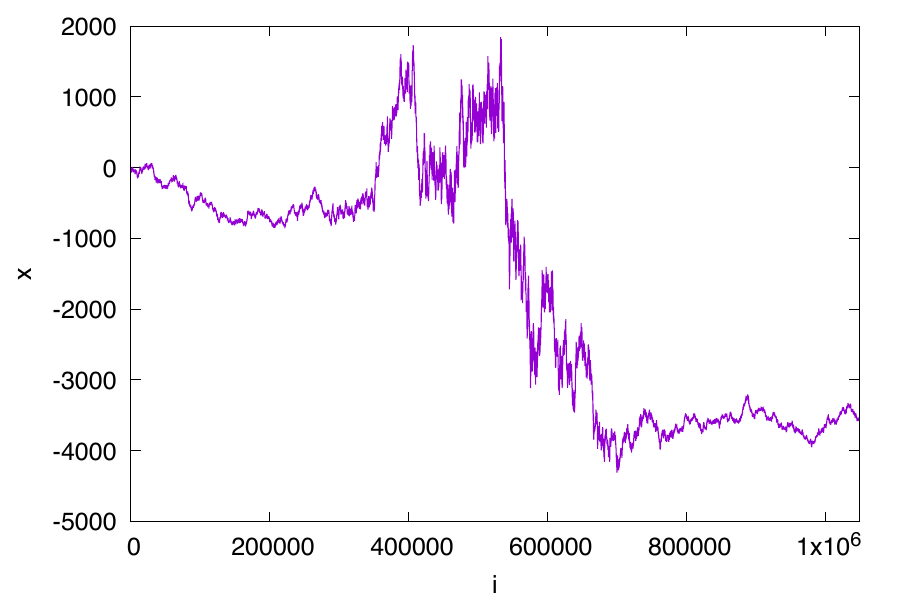} 
\caption{\label{fig:arn} A realization of the $AR(N)$ process described by equation (\ref{arN}) with $N=2^{20}$, $\alpha = 2$ and $\beta = 9$.}
\end{figure}

\begin{figure}[htp]
\includegraphics[width=0.48\textwidth]{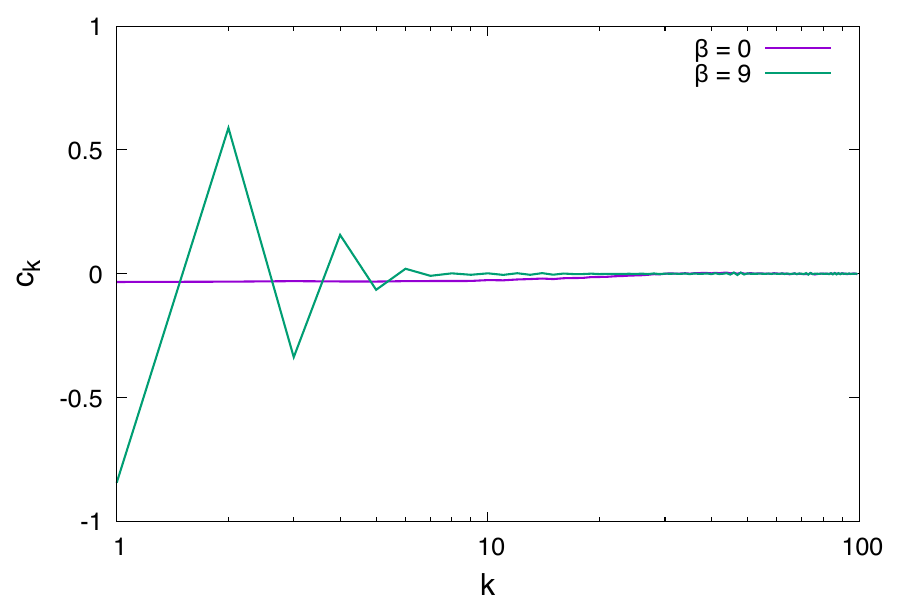} 
\caption{\label{fig:arncorr}  Fourier domain correlation coefficients $c_k$ for the $AR(N)$ process with $\alpha = 2$ and $\beta = 0$ (stationary case) and  $\beta = 9$ (non-stationary). The index $k$ corresponds to the number of frequency bins between the data samples used to compute the correlation.}
\end{figure}

As discussed earlier, a major advantage of the WDM wavelet basis over a Fourier basis is that it yields a diagonal noise correlation for a range of non-stationary processes. As a concrete example of a non-stationary process with an evolutionary spectra, consider the $AR(N)$ model with time varying variance:
\begin{equation}\label{arN}
\sum_{j=0}^N a_j x[i-n] = \left(1+ \beta e^{-\frac{25 (i-N/2)^2}{ 2 N^2}}\right) \delta_i \, ,
\end{equation}
where $\delta_i$ is a zero norm, unit variance Gaussian random variate, $\delta_i \sim {\cal N}(0,1)$, and $N$ is the total number of time samples $x[i]$. Setting the coefficients $a_j$ such that
\begin{equation}
a_j = \left(\frac{2j -2 - \alpha}{2j} \right) a_{j-1}
\end{equation}
with $a_0=1$ generates colored noise with a $1/f^\alpha$ spectrum~\cite{381848}. When $\beta$ is non-zero the process is non-stationary. Figure~\ref{fig:arn}
shows one realization of this $AR(N)$ process with $N=2^{20}$, $\alpha = 2$ and $\beta = 9$. The non-stationarity is evident in the increased variance of the process for samples near the midpoint. The non-stationarity can also be seen in the Fourier domain correlation coefficients
$c_k=C_k/C_0$, where
\begin{equation}
C_k =\frac{1}{N-k}  \sum_{i=0}^{N-1 - k}  \left( \bar{X}[i]  \bar{X}^*[i+k] +  \bar{X}^*[i]  \bar{X}[i+k] \right) \, ,
\end{equation}
and $\bar{X}[i]$ are the whitened Fourier coefficients. Here these are found by taking a FFT of the time domain data and dividing the Fourier coefficients by the square root of the theoretical $1/f^\alpha$ spectrum. To reduce spectral leakage, a Tukey filter that is flat across 95\% of the data was applied to the time domain data before transforming to the frequency domain. Figure~\ref{fig:arncorr} shows the first one hundred Fourier domain correlation coefficients for stationary ($\beta =0$) and non-stationary ($\beta=9$) realizations of the $AR(N)$ process. Strong correlations between Fourier bins are evident for non-stationary case. The correlation coefficients for the stationary case deviate slightly from zero due to the application of the Tukey window. As discussed in more detail later, apodizing windows introduce non-stationarity to otherwise stationary time series.

The evolutionary spectrum $S(t,f)$ can be estimated in many different ways. One method is to introduce a parametric model for the evolutionary spectra, $S_\theta(t,f)$, and solve for the parameters of the model, $\theta$. Under the assumption that the noise is Gaussian distributed, the model parameters can be estimated by maximizing the Whittle log likelihood
\begin{equation}
\ln p(d| \theta) = - \frac{1}{2} \sum_{n,m} \left( \ln ( 2\pi S_\theta(t_n, f_m)) + \frac{w_{nm}^2}{S_\theta(t_n, f_m)} \right) \, .
\end{equation}
Alternatively, the Whittle likelihood can be used to infer the model parameters in a Bayesian setting. To avoid over-parameterizing the model, trans-dimensional methods are preferable. For example, the {\tt Bayesline} spectral estimation algorithm~\cite{PhysRevD.91.084034} can be generalized to use a bicubic spline model, where the number of spline points and their amplitudes are allowed to vary. A much simpler, though less accurate estimate for the evolutionary spectra can be found using a fixed grid of bicubic spline points and estimating their amplitude from the median of the points in the scalogram, $w_{nm}^2$, surrounding the spline control point. The median is more robust to outliers than the mean. The median of the scalogram has to be divided by a factor of ${\tilde \chi}^2_1/\bar{\chi}^2_1$ to yield an estimate of $S(t,f)$ under the assumption that the scalogram is chi-squared distributed with one degree of freedom. Here $\bar{\chi}^2_1=1$  and ${\tilde \chi}^2_1/\bar{\chi}^2_1=0.4549364231...$ are the mean and median of the chi-squared distribution. The median is found from the transcendental equation $\Gamma[\frac{1}{2}, \frac{1}{2} {\tilde \chi}^2_1] = \frac{1}{2} \Gamma[\frac{1}{2}]$, where $\Gamma[a,b]$ and $\Gamma[a]$ are the incomplete and complete Gamma functions, respectively.

\begin{figure}[htp]
\includegraphics[width=0.48\textwidth]{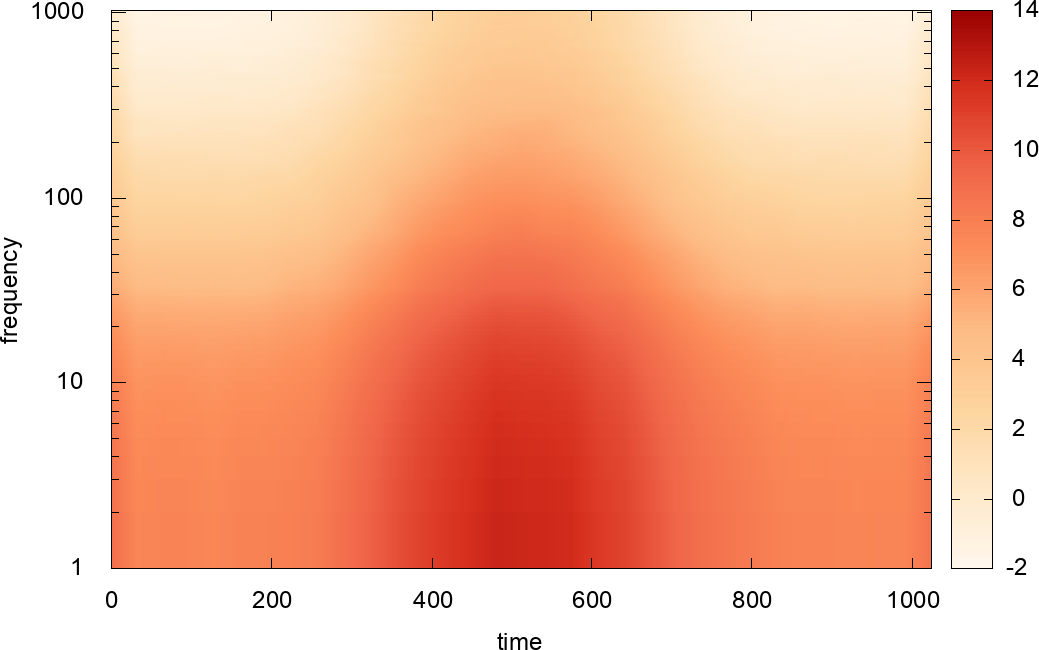} 
\caption{\label{fig:sft} A bicubic spline fit to logarithm of the evolutionary spectrum, $\ln S(f,t)$, for the non-stationary $AR(N)$ model described by equation (\ref{arN}) with $\alpha = 2$ and $\beta = 9$.}
\end{figure}

Figure~\ref{fig:sft} shows a bicubic spline fit to the evolutionary spectrum for the $AR(N)$ process (\ref{arN}) with $\alpha=2$ and $\beta=9$. The spectrum varies over seven decades in magnitude. The time series was generated with $N=2^{20}$ points and the WDM transform was computed on a grid with $N_t=N_f=1024$ points. The bicubic spline control points were spaced at intervals of 32 points in the both time and frequency directions. The whitened wavelet amplitudes, $W_{nm} = w_{nm}/\sqrt{S_\theta(t_n, f_m)}$ follow a ${\cal N}(0,1)$ unit normal distribution (verified using an Anderson-Darling test). 

\begin{figure}[htp]
\includegraphics[width=0.48\textwidth]{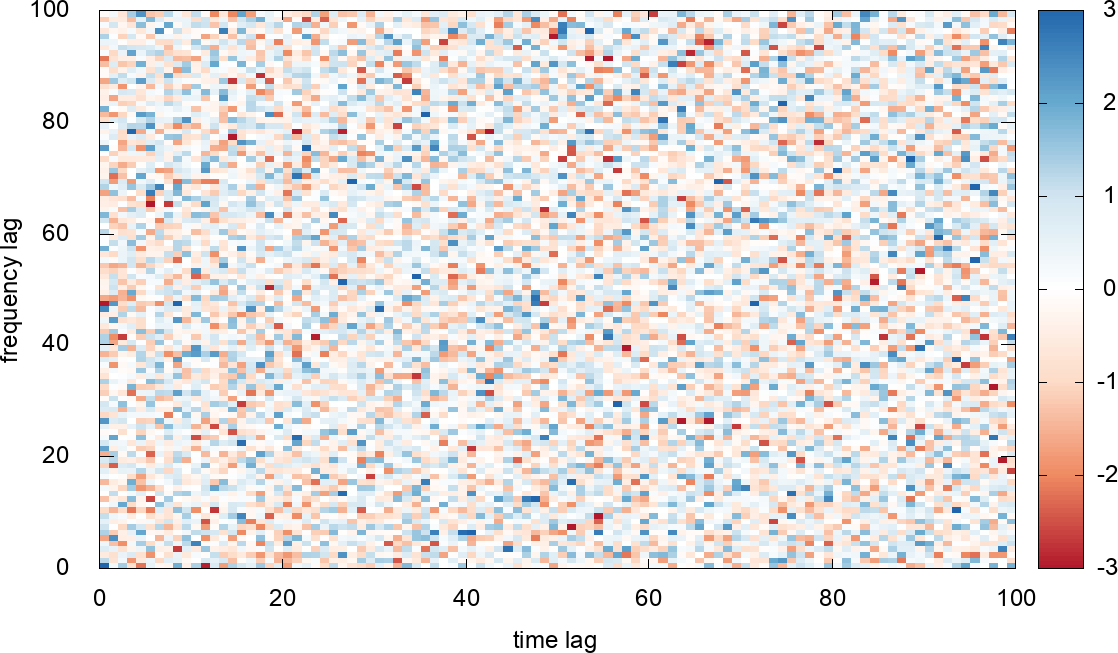} 
\caption{\label{fig:corr} The scaled off-diagonal noise correlation coefficients $\bar{c}_{ij}$ of the WDM wavelet transform for the non-stationary $AR(N)$ model described by equation (\ref{arN}) with $\alpha = 2$ and $\beta = 9$.}
\end{figure}

\begin{figure}[htp]
\includegraphics[width=0.48\textwidth]{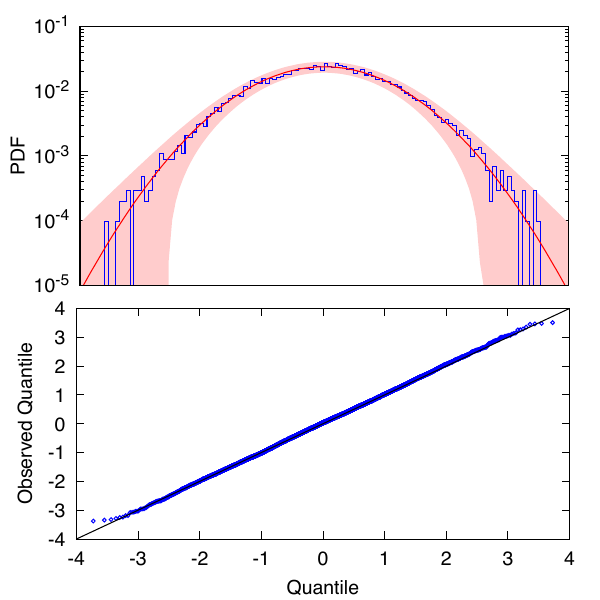} 
\caption{\label{fig:ncorr} Histogram (upper panel) and quantile-quantile plot (lower panel) for the scaled off-diagonal noise correlation coefficients $\bar{c}_{ij}$ shown in Figure~\ref{fig:corr}. The $\bar{c}_{ij}$ follow a ${\cal N}(0,1)$ unit normal distribution, confirming that the WDM noise correlation matrix is diagonal.}
\end{figure}

Under the hypothesis that the wavelet noise correlation matrix is diagonal, the products $W_{nm}W_{n' m'}$ for $n \neq n'$ and/or $m \neq m'$ should follow a product normal distribution, $p(x) = K_0(|x|)/\pi$, where $K_0(x)$ is a modified Bessel function of the second kind. A histogram of the products confirms this expectation. A more quantitative test is to look at estimates of the whitened noise correlation matrix
\begin{equation}
c_{ij} = {\rm E}[W_{nm}W_{n+i, m+j}] \approx \frac{1}{N} \sum_{n,m} W_{nm} W_{n+i, m+j} \, .
\end{equation}
In the limit that $N \rightarrow \infty$, the off-diagonal elements of $c_{ij}$ should vanish. To allow for the finite number of samples, the $c_{ij}$ can be scaled by the standard deviation of the sample mean, which is given by $\sigma^2 = {\rm Var}[W_{nm}W_{n+i, m+j}]/N$. The scaled correlations, $\bar{c}_{ij} = c_{ij}/\sigma$ should follow a ${\cal N}(0,1)$ unit normal distribution for $i \neq j$. Figure~\ref{fig:corr} shows the first ten thousand $\bar{c}_{ij}$ for the data shown in Figure~\ref{fig:arn}. The distribution appears to be consistent with unit variance white noise, with no discernible correlation pattern. Figure~\ref{fig:ncorr} shows a histogram and quantile-quantile plot of the coefficients, and confirms that they follow a ${\cal N}(0,1)$ unit normal distribution.

Similar tests were applied to a variety of noise processes, including stationary white noise, stationary red noise, cyclo-stationary noise, non-stationary white noise with variance that increased linearly with time, and non-stationary red noise generated by the $AR(N)$ process with different values of $\alpha$ and $\beta$. In all cases the WDM wavelet noise correlation matrix was found to be diagonal. 

 \begin{figure}[htp]
\includegraphics[width=0.48\textwidth]{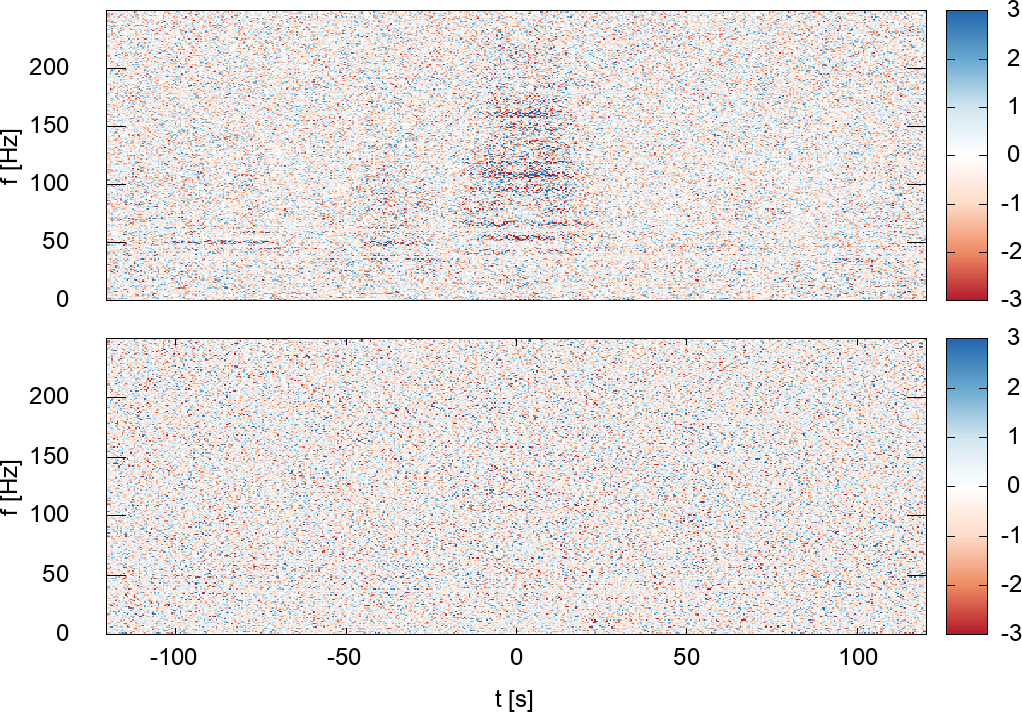} 
\caption{\label{fig:Hanford} The upper panel shows the WDT transform of 256 seconds of data from the LIGO Hanford detector centered on GPS time 1165067917. The data has been whitened using the the median wavelet PSD $S(f)$. The lower panel shows the same data whitened and made stationary by the evolutionary spectrum PSD $S(f,t)$.}
\end{figure}

\begin{figure}[htp]
\includegraphics[width=0.48\textwidth]{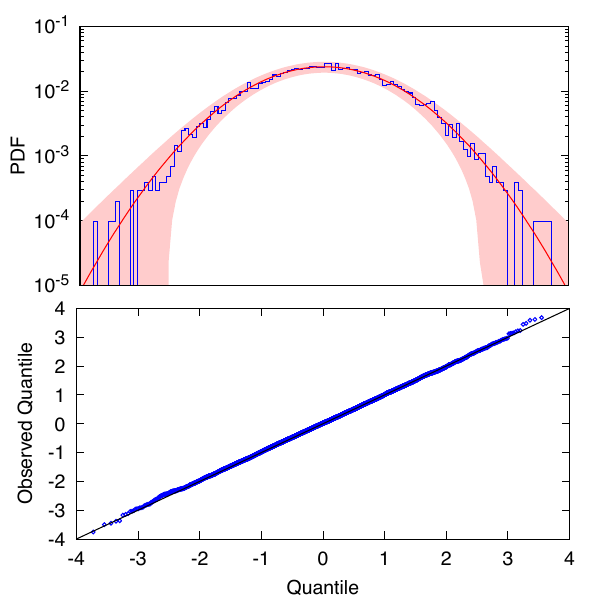} 
\caption{\label{fig:Hcorr} Histogram (upper panel) and quantile-quantile plot (lower panel) for the scaled off-diagonal noise correlation coefficients $\bar{c}_{ij}$ for the LIGO Hanford data shown in Figure~\ref{fig:Hanford}. The $\bar{c}_{ij}$ follow a ${\cal N}(0,1)$ unit normal distribution, confirming that the WDM noise correlation matrix is diagonal.}
\end{figure}

A more relevant question is whether the WDM transform yields a diagonal noise correlation matrix when applied to real gravitational wave data. The LIGO--Virgo data analysis guide paper~\cite{LIGOScientific:2019hgc} investigated a particular stretch of LIGO Hanford data that exhibited non-stationary noise. The WDM scalogram of the same 256 second stretch of data studied in Ref.~\cite{LIGOScientific:2019hgc} is shown in Figure~\ref{fig:Hanford}, focusing on the frequency band below 250 Hz where the non-stationarity is most pronounced. The WDM transform had a time resolution of $\Delta T=0.5$ s and a frequency resolution of $\Delta F = 1$ Hz. The upper panel in Figure~\ref{fig:Hanford} shows the whitened data, using an estimate for $S(f)$ based on the median of the scalogram in each frequency band. The lower panel in Figure~\ref{fig:Hanford} shows the same data, now whitened with the evolutionary spectrum $S(f,t)$. Applying the same analysis to the off-diagonal noise correlation coefficients, $c_{ij}$, as was done previously for the non-stationary $AR(N)$ model yields the distribution shown in Figure~\ref{fig:Hcorr}. The scaled coefficients $\bar{c}_{ij}$ follow a ${\cal N}(0,1)$ unit normal distribution, confirming that the WDM noise correlation matrix for this stretch of non-stationary LIGO Hanford data is diagonal.

\subsection{Gaps and Edge effects}

\begin{figure}[htp]
\includegraphics[width=0.48\textwidth]{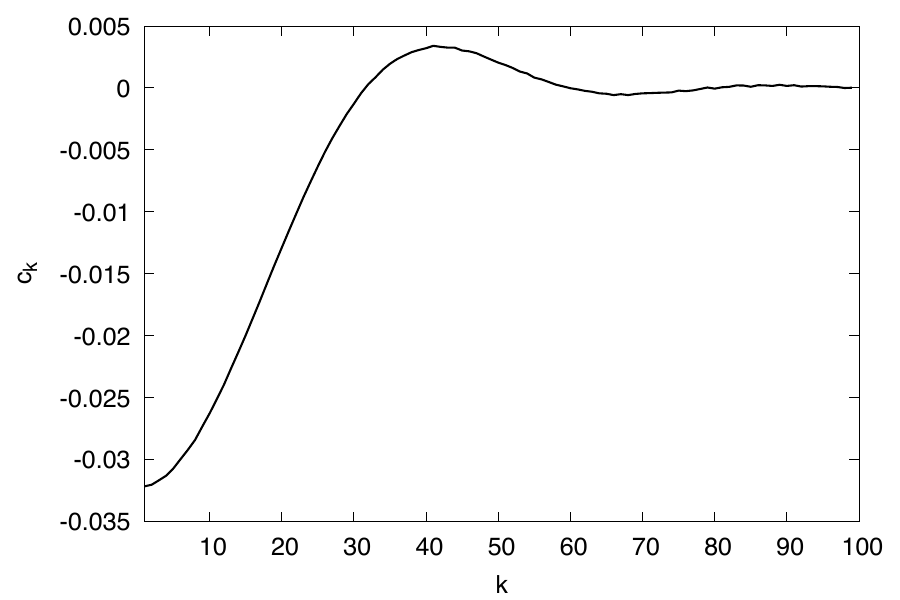} 
\caption{\label{fig:fcorr} The off diagonal elements of the Fourier domain noise correlation function $c_k$ for stationary white noise with a Tukey window function applied. The index $k$ corresponds to the number of frequency bins between the data samples used to compute the correlation.}
\end{figure}

It is well known that time domain data with edges and gaps lead to spectral leakage in the Fourier domain. The spectral leakage can be mitigated using window functions, but at the cost of lost data, and artificial non-stationarity. The latter point is not widely appreciated. Even using a Tukey window that is flat across a large fraction on the data results in a non-diagonal noise correlation matrix in the Fourier domain. Figure~\ref{fig:fcorr} shows the Fourier domain noise correlation function $c_k$ for 16 seconds of stationary white noise, $x[i]$, sampled at 4096 Hz, with a Tukey window applied. The Tukey window had a rise and decay time of 0.4 s to mimic the settings in typical LIGO analyses~\cite{LIGOScientific:2019hgc}. The window is flat across $1-\alpha$ of the data, with $\alpha = 0.8/16= 0.05$. The magnitude of the maximum correlation is of order $\alpha$. 

\begin{figure}[htp]
\includegraphics[width=0.48\textwidth]{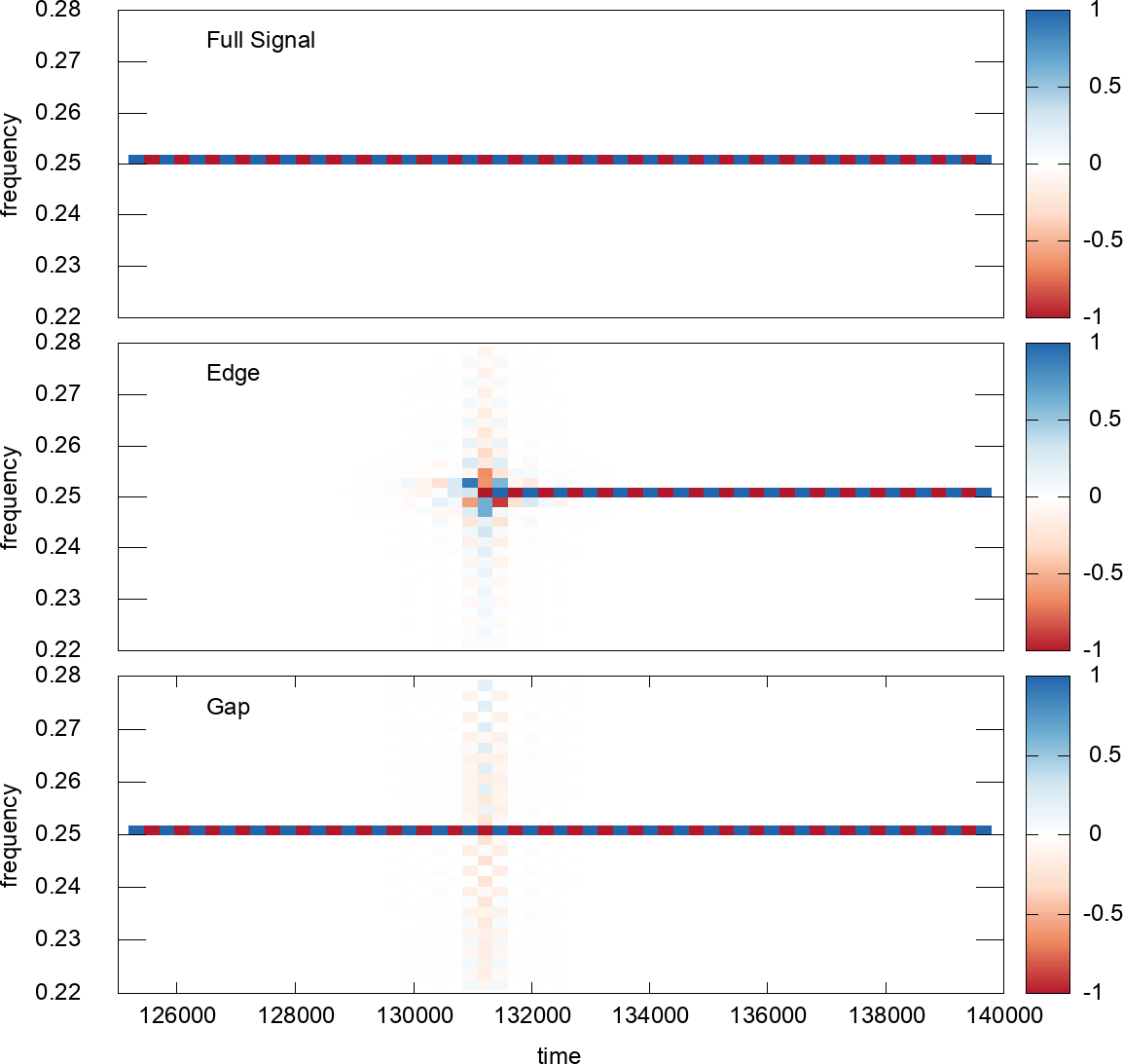} 
\caption{\label{fig:cosine} The WDM transform of a sine wave (upper panel), a sine wave with an edge at time 131072 (middle panel), and a sine wave with a gap of width $\Delta T = 1/32$ (lower panel). The amplitude of the sine wave is 8, but the color scale has been capped between $[-1,1]$ to better show the leakage of power caused by the edge and gap.}
\end{figure}

Wavelet transforms are not immune to the edge effects and gaps, though the impact is more localized. Edges and gaps introduce spurious sharp features in the data that cause Gibbs oscillation and spectral leakage, as also happens with a Fourier transform. Another way of looking at it is that the edges and gaps render the wavelet basis functions non-orthogonal, resulting in leakage of power between wavelet pixels. The leakage is confined to wavelets whose time domain filters overlap with the edge or gap. Recall that the time domain window functions are always wider than the wavelet pixels time span $\Delta T$, typically by factors of order ten. Figure~\ref{fig:cosine} illustrates the effects of edges and short duration gaps (less than the width of wavelet pixel) on the WDM transform of a sine wave. The time domain window function for the WDM transform had width $q \Delta T$ with $q=16$. The power leakage due to the signal having a sharp edge extends for order $16 \Delta T$ in time and  $32 \Delta F$ in frequency. The short data gap leads to a similar pattern of leakage, with a slightly smaller extent in time and slightly larger extent in frequency. 

Methods for mitigating the effects of edges and gaps that are used in frequency domain analyses, such as apodizing the edges with a window function, and filling the gaps with fake data using Bayesian data augmentation~\cite{Baghi:2019eqo} or otherwise~\cite{Zackay:2019kkv}, can also be used with wavelet transforms. An alternative is to introduce specialized boundary wavelets and gap wavelets that maintain orthogonality and eliminate leakage~\cite{Andersson_waveletson,cohen:hal-01311753,Williams95adiscrete}, but at the cost of added computational complexity, and a decrease in frequency separation in the regions covered by these special wavelets.

\subsection{Glitches and spectral lines}

\begin{figure}[htp]
\includegraphics[width=0.48\textwidth]{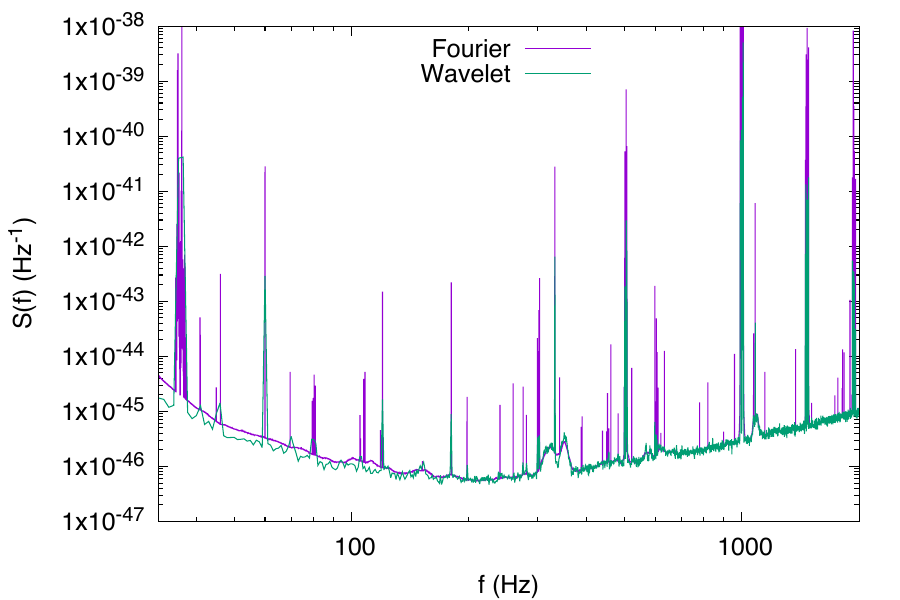} 
\caption{\label{fig:PSD} Estimates of the power spectral density of the LIGO Hanford data shown in Figure~\ref{fig:Hanford} using Fourier domain and wavelet domain methods. The spectrum exhibits many sharp spectral lines. Some of the weaker lines are smoothed away in the wavelet domain analysis.}
\end{figure}

The data from ground based interferometers exhibit sharp spectral features and suffer from frequent noise transients, or glitches~\cite{LIGOScientific:2019hgc}. Figure~\ref{fig:PSD} compares Fourier domain and wavelet domain estimates of the power spectral density of the 256 seconds of LIGO Hanford data used to produce 
Figure~\ref{fig:Hanford}. The Fourier domain estimate used a running median of the Fourier periodogram, combined with an outlier identification to catch the spectral lines. The wavelet domain estimate used the median of the wavelet scalogram across each frequency band. The spectrum contains numerous spectral lines. The weaker lines are absent in the wavelet estimate as the contribution from the weaker lines is a small contribution to the total integrated power across a wavelet. The wavelet estimate is more robust against the non-stationary excess below 250 Hz, which is why the wavelet estimate sits below the Fourier domain estimate at low frequencies.

Spectral lines can be both very narrow in frequency and very large in amplitude. Strong lines can saturate an entire wavelet frequency band, rending the data useless in that band. The fraction of otherwise useable data lost to strong lines scales as the ratio of the width of the line $\delta f$ to the bandwidth of the wavelet layer, $\Delta F$. This suggests choosing $\Delta F$ small to avoid losing information. However, small $\Delta F$ implies large $\Delta T$, and transforms with large $\Delta T$ are less effective in dealing with non-stationarity. A way around this problem is to first transform the data using a small value for $\Delta F$, and use this transform to estimate the power spectrum $S(f)$, ignoring any non-stationarity for now. The data can then be whitened using this estimate, returned to the time or frequency domain, then wavelet transforming a second time using a larger $\Delta F$. The evolutionary power spectrum of the re-transformed data, $\bar{S}(t,f)$, can then be computed. The full dynamic power spectrum is given by $S(f,t) = S(f) \bar{S}(t,f)$.

\begin{figure}[htp]
\includegraphics[width=0.48\textwidth]{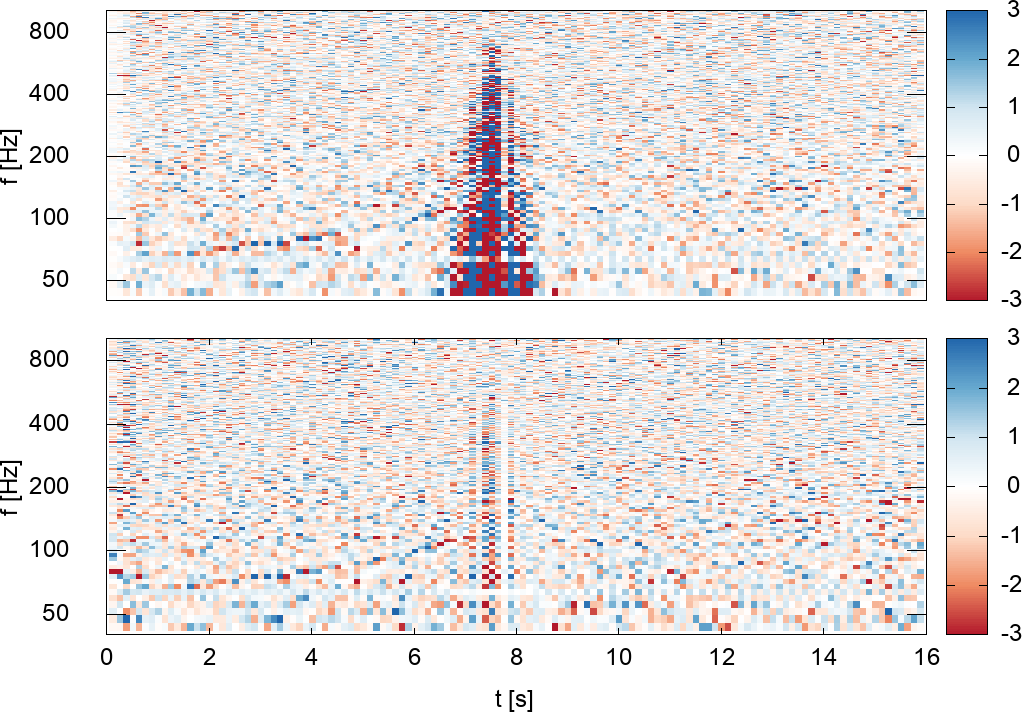} 
\caption{\label{fig:GW170817} The upper panel shows the WDT transform of 16 seconds of data from the LIGO Livingston detector surrounding the BNS merger GW170817. The data has been whitened using the the median wavelet PSD $S(f)$. The lower panel shows the same data whitened and made stationary by the evolutionary spectrum $S(f,t)$. The loud noise transient is significantly reduced by the evolutionary PSD. The inspiral track of the GW signal is visible in lower left of both panels.}
\end{figure}

In addition to slow drifts in the evolutionary power spectrum, such as those seen in the LIGO Hanford data in Figure~\ref{fig:Hanford}, interferometer data also manifest frequent short duration noise transients, or glitches. In contrast to long term drifts, which can be modeled as Gaussian noise with a time varying power spectrum, glitches are short duration, coherently structured non-Gaussian features. A better term for them might be ``instrument generated signals''. Figure~\ref{fig:GW170817} shows the WDM transform of 16 seconds of LIGO Livingston data surrounding the binary neutron star merger GW170817~\cite{TheLIGOScientific:2017qsa}. A very loud glitch can be seen roughly one second prior to merger. Whitening the data with the evolutionary wavelet spectrum helps mask the effect of the glitch, but a much better approach is to coherently model and remove the glitch from the data, as was done in the original analysis of GW170817~\cite{TheLIGOScientific:2017qsa}, where the {\tt BayesWave} algorithm~\cite{Cornish:2014kda} was used to subtract the glitch from the data. Treating glitches as instrument signals and removing them from the data is superior to gating the affected data or whitening the data with the evolutionary spectrum, as the later approaches lead to a loss of information, while the former approach preserves all of the data~\cite{Pankow:2018qpo}.

\section{Wavelet Domain Waveforms}

Gravitational wave data analysis often requires millions of likelihood calculations, each involving one or more waveform evaluations. While the computational cost of the WDM wavelet transform is comparable to a standard FFT based frequency domain analysis, the cost can be prohibitive for long-lived signals. To reduce the computational cost of standard frequency domain analyses, significant effort has gone into producing waveform models that can be evaluated directly in the frequency domain~\cite{Buonanno:2009zt,PhysRevD.77.104017,PhysRevLett.106.241101,Santamaria:2010yb,Hannam:2013oca,Khan:2015jqa,Garcia-Quiros:2020qpx,Chatziioannou:2017tdw,Moore:2018kvz}. The goal here is to develop fast wavelet domain waveforms using time domain or frequency domain waveforms as a starting point. Remarkably, it turns out that waveforms can be generated more efficiently in the wavelet domain: the computational cost scales as the square root of the number of data points, as opposed to linearly, or worse, for the time and frequency domains. 

Let us assume that the waveform templates for binary systems can be expressed in the form
\begin{equation}\label{tsum}
h(t) = \Re \sum_{k} A_k(t) e^{i \Psi_k(t)} \, ,
\end{equation}
or equivalently
\begin{equation}\label{fsum}
{\tilde h}(f) = \sum_{k} {\cal A}_k(f) e^{i \Theta_k(f)} \, .
\end{equation}
Here the amplitude and phase are taken to include contributions from the instrument response in addition to contributions from the evolution of the binary system. The series does not have to be harmonic, though that will often be the case~\cite{Roy:2019phx}. The fast wavelet transform is applied term by term in the sums.
 In some cases only a few terms are needed to accurately model the signal, such as for comparable mass spin-aligned binaries on quasi-circular orbits. In other cases a large number of terms will be needed, such as for system on highly eccentric orbits with large mass ratios. The efficiency of the rapid wavelet transforms degrades as the number of terms in the sum increases, eventually reaching the point where there are no savings to be gained over directly transforming the full signal. Similar considerations apply when computing frequency domain waveforms for eccentric binaries~\cite{Moore:2018kvz}.
 
 \begin{figure}[htp]
\includegraphics[width=0.5\textwidth]{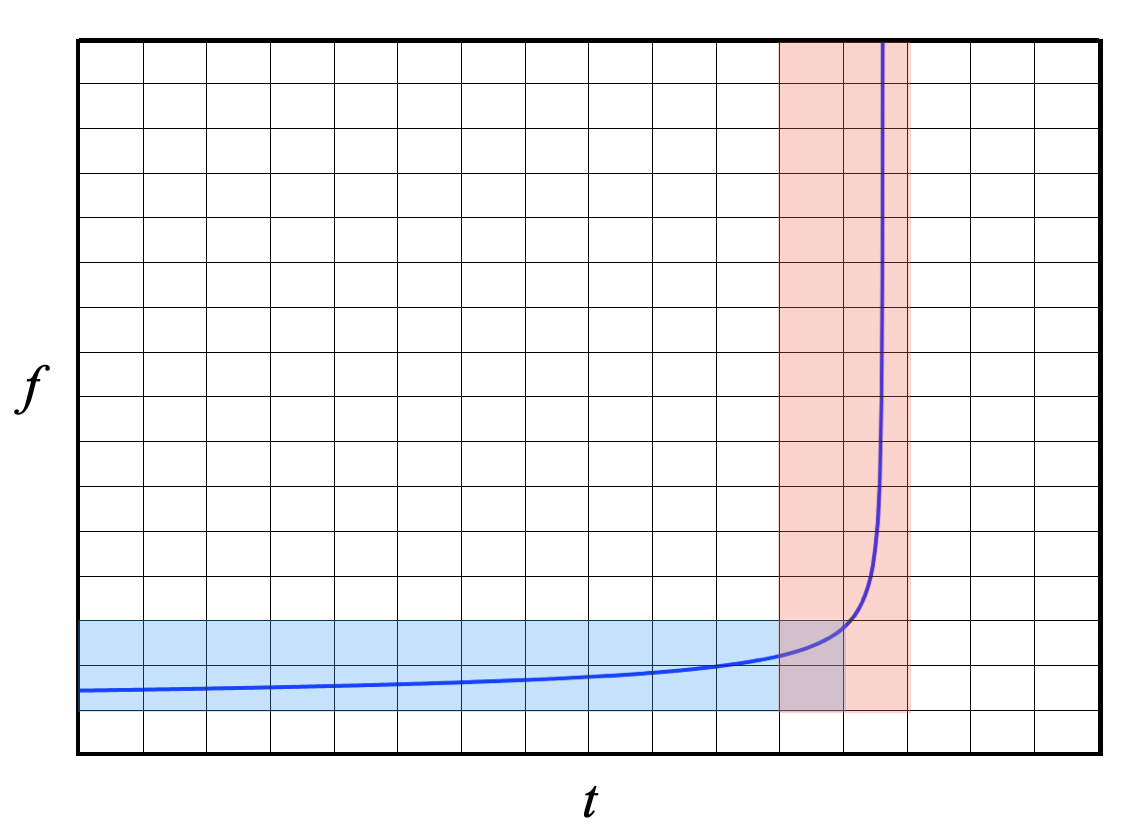} 
\caption{\label{fig:track} An illustration of the time-frequency decomposition of a single term in the expansion of the signal from a binary merger. The horizontal (light blue) shaded region indicates where fast time-domain transforms are most efficient, while the vertical (light red) shaded region indicates where fast frequency-domain transforms are most efficient.}
\end{figure}
 
 Figure~\ref{fig:track} illustrates the time-frequency evolution of one term in the sums (\ref{tsum}), (\ref{fsum}). During the early inspiral (horizontal shaded region) the signal evolves slowly in time, and fast time-domain transforms are most efficient. During the late inspiral, merger and ringdown (vertical shaded region) the signal evolves rapidly in frequency, and fast-frequency domain transforms are most efficient. In each region there are (at least) two fast methods for computing the wavelet transform. The first method, sparse sampling, involves no approximations and is typically one or two orders of magnitude faster than the direct transform. The second method employs a local Taylor series expansion of the amplitude and phase, and interpolation over a look-up table. The second method is typically two to three orders of magnitude faster than the direct transform. The computational savings are greatest for long-lived signals, such as binary Neutron star mergers for ground based detectors and systems with masses below $10^5 M_\odot$ for space based detectors.
 
 \subsection{Fast Wavelet Waveforms: Sparse Sampling}
 
 The sparse sampling technique employs a stationary-phase inspired time-frequency mapping defined:
 \begin{eqnarray}\label{spa1}
 t_k(f) &=& \frac{1}{2\pi} \frac{d \Theta_k(f)}{df} \nonumber \\
 f_k(t) &=& \frac{1}{2\pi} \frac{d \Psi_k(t)}{dt}  \, ,
 \end{eqnarray}
 and similarly for the time and frequency derivatives:
 \begin{eqnarray}\label{spa2}
 t'_k(f) &=& \frac{1}{2\pi} \frac{d^2 \Theta_k(f)}{d^2f} \nonumber \\
{\dot f}_k(t) &=& \frac{1}{2\pi} \frac{d^2 \Psi_k(t)}{d^2t}  \, .
 \end{eqnarray}
 The sparse sampling approach can be understood as a form of heterodyning, or equivalently, a stationary phase evaluation. At time $t_k = n \Delta T$ the signal has a frequency $f_k \approx m \Delta F$ for some specific $m$. Alternatively, the signal has frequency $f_k = m \Delta F$ at some time $t_k \approx n \Delta T$ for some specific $n$. At these instances, the oscillatory factors in equations (\ref{fulltx}) and (\ref{fullfx}) largely cancel the oscillations due to the phase terms $e^{i \Psi_k(t)}$ or $e^{i \Theta_k(f)}$. Since the summands are then slowly varying, the sums can be evaluated using a much sparser sampling in time or frequency, thereby significantly  reducing the computational cost. Time-domain sparse sampling is most efficient in regions where the signal is slowly evolving in frequency:  $\dot f_k T_{\rm w} < \Delta F$, where $T_{\rm w}$ is the duration of the wavelet window function.  Frequency-domain sampling is most efficient in regions where the signal is rapidly evolving in frequency: $t'_k F_{\rm w} < \Delta T$, where $F_{\rm w}$ is the bandwidth of the wavelet window function. 
 
Sparse sampling in the time domain works by reducing the sample cadence such that the time samples are spaced by $p \, \Delta t$, where $p$ is some power of two. The larger the $p$ the greater the savings in computational cost. The downsampling is limited by how rapidly the amplitude and heterodyned phase vary. Significant savings ($p=256$ or larger) are only possible when the amplitude and heterodyned phase are evolving relatively slowly in time. Considering a single harmonic of the waveform, and dropping the subscripts to simplify the notation, the wavelet coefficients are given by
\begin{equation} 
w_{nm} =  \sqrt{2} \Delta t \Re\, C_{nm} S_{n}[(m-m_c)q] \, .
\end{equation}
where $m_c = [f(n \Delta T)/\Delta F]$ is the central frequency band and
\begin{equation}\label{sparseT}
S_{n}[l] =  \sum_{j = -L/2}^{L/2-1}  e^{2\pi i j l/L} A[nN_f/p] \phi[j]  e^{i(\Psi[n N_f/p] - 2\pi m_c q j /L)}
\end{equation}
with $L = K/p$. The $S_{nl}$ can be computed using a FFT at cost $L \ln L$ for each $n$, yielding a total cost that scales as $N_t L \ln(L)$. This is more than factor of $p$ faster than the full time domain transform. The sparse transform is exact so long as the summand in (\ref{sparseT}) is adequately sampled. The downsampling factor $p$ can be dynamically adjusted according to how fast the signal is evolving.

Sparse sampling in the frequency domain works by decreasing the time span $T \rightarrow T/r$ such that the frequency samples are spaced by $r/T$, where $r$ is some power of two. The time span must be equal to or greater than the window size $T_w = K \Delta t$, so that $r \leq N/K$. Since the frequency domain method is most useful when the signals are rapidly evolving, we saturate the bound and set $r=N_t/(2q)$. The wavelet coefficients are then given by
\begin{equation}
w_{nm} =  \sqrt{2} (-1)^{nm}  \Re\, C_{nm}  s_m[n-n_c]\, ,
\end{equation}
where $n_c = [t(m \Delta F)/\Delta T]$ is the central time slice and 
\begin{equation}\label{sparseF}
s_{m}[j] = \sum_{l = -q}^{q-1}    e^{-\pi i l j/q} {\cal A}[l+mq] \Phi[l]  e^{i(\Theta[l+mq] + \pi l n_c /q)} \, .
\end{equation}
Evaluating (\ref{sparseF}) for each $m$ using a FFT yields a total cost of $2 N_f q \ln(2q)$, which is a factor of $r=N_t/(2q)$ less than the full frequency domain transform. The limitation of using a time span equal to the window size is that the time is periodic with period $T_w$, meaning that the transformation becomes multi-valued for slowly evolving signals. For (\ref{sparseF}) to be single valued we require that the signal evolves to occupy a new frequency layer in the time it takes to cross the window function. That is, we require $\dot f \, T_{\rm w} > \Delta F$. When this condition is met the sparse frequency domain transform is exact.

 \subsection{Fast Wavelet Waveforms: Taylor expansion}
 
\begin{figure}[htp]
\includegraphics[width=0.5\textwidth]{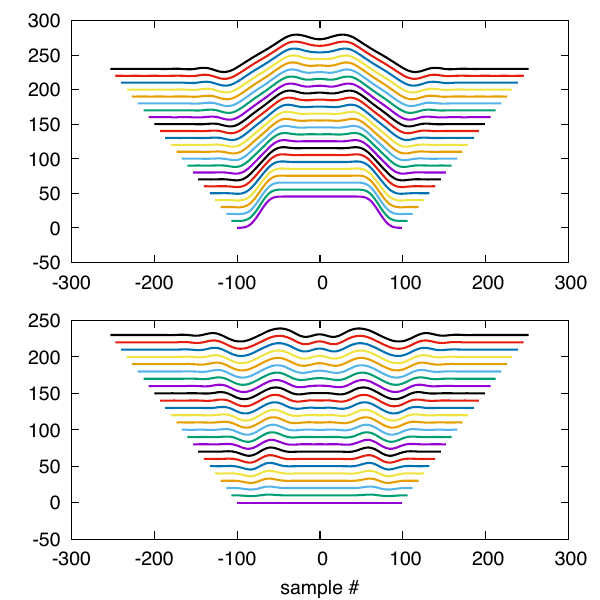} 
\caption{\label{fig:taylort} An illustration of the lookup table for the cosine (upper) and sine (lower) coefficients used to compute the time domain expansion coefficients $c_{nm}(f,\dot f)$ and $s_{nm}(f,\dot f)$. The horizontal axis shows the successive frequency samples, while the vertically offset lines show increments of the frequency derivative.}
\end{figure}

 The second fast transform takes advantage of the fact that the individual harmonics of the gravitational wave signal from a binary system go from evolving slowly in frequency to evolving slowly in time (see Figure~\ref{fig:track}). During the early inspiral the signals evolve slowly in time, and the amplitude are frequency are
 approximately constant across an individual wavelet.  Consider a wavelet centered at $t_n= n \Delta T$. The gravitational wave amplitude and phase can be Taylor expanded:
 \begin{eqnarray}
  \Psi(t) &=&   \Psi(t_n) + 2 \pi (t-t_n)  f(t_n) + \pi (t-t_n)^2 \dot f(t_n)  + \dots \nonumber \\
 A(t) &= &A(t_n) + (t-t_n) \dot A(t_n)  + \dots 
 \end{eqnarray}
 where the expressions for $f(t)$ and $\dot f(t)$ are given in equations (\ref{spa1}) and (\ref{spa2}). To achieve high accuracy these expansions can be continued to higher order, but it is usually enough to stop at second order in the phase and zeroth order in the amplitude. The wavelet coefficients
 \begin{equation}
 w_{nm} =  \sum_{k = 1}^{N} g_{nm}[k] A(t) e^{i \Psi(t)}
\end{equation}
 are then given by
 \begin{eqnarray}
 &&w_{nm} = A(t_n) \left( c_{nm}(f(t_n),\dot f(t_n)) \cos \Psi(t_n) \right. \nonumber \\
&& \hspace*{0.8in} \left. -  s_{nm}(f(t_n),\dot f(t_n)) \sin\Psi(t_n)\right)
\end{eqnarray}
where
\begin{equation}
 c_{nm}(f,\dot f) =  \int dt \, g_{nm}(t) \cos(2 \pi (t-t_n)  f + \pi (t-t_n)^2 \dot f) 
\end{equation}
and similarly for $s_{nm}(f,\dot f)$ with the cosine replaced by a sine. The $c_{nm}$ and $s_{nm}$ can be pre-computed on a grid in $(f,\dot f)$ and interpolated. This procedure can be used for any wavelet family. The WDM wavelets offer an additional advantage in that they have a uniform window in frequency, so that the coefficients, computed relative to the central frequency of one layer, are the same for all layers. This dramatically shrinks the size of the lookup table since it only has to be computed for a single frequency layer. The lookup table is generated on a grid spaced by $\delta f = \epsilon \Delta F$ with $\epsilon \approx 0.01$ and $\delta \dot f = \varepsilon \Delta F/T_w$ with $\varepsilon \simeq 0.1$. The number of frequency derivative samples depends on the maximum frequency we wish to cover - typically $\dot f_{\rm max} = 8 \Delta F/T_w$ is sufficient. The number of frequency samples required depends on the filter bandwidth $(A+B)/\pi$ and the frequency evolution of the signal, which increases with increasing $\dot f$. The total number of frequency samples scales as $((A+B)/\pi + \dot f_{\rm max} T_w)/\delta f$. The full lookup table typically has order $10^5$ entries. For very slowly evolving signals, such as LISA galactic binaries, the number is far less - of order 100. Note that since $((A+B)/\pi + \dot f T_w) > \Delta F$, the wavelet coefficients at each time sample $n$ often span several frequency layers $m$.
Figure~\ref{fig:taylort} illustrates the time domain expansion coefficients $c_{nm}(f,\dot f)$, $s_{nm}(f,\dot f)$ as a function of the instantaneous frequency (across the horizontal) and frequency derivative (vertically offset lines) of the signal. The WDM wavelets used to generate this figure used the same settings as those used in Figure~\ref{fig:window}, namely $d=4$, $A = \Delta \Omega/4$, $B = \Delta \Omega/2$ and $q=16$.

\begin{figure}[htp]
\includegraphics[width=0.5\textwidth]{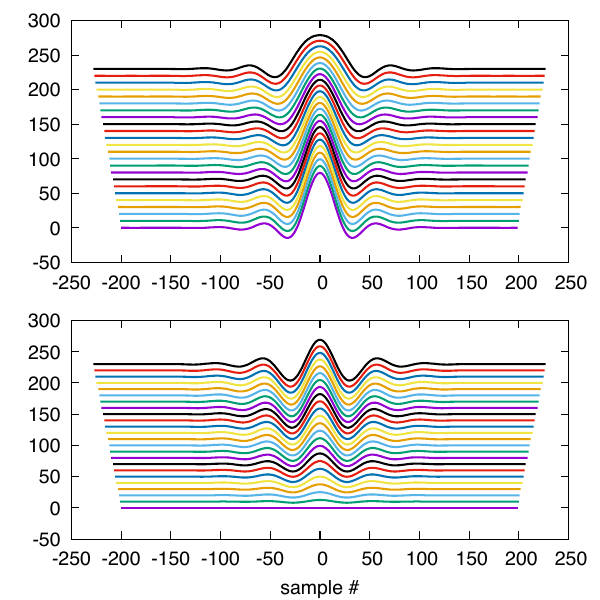} 
\caption{\label{fig:taylorf} An illustration of the lookup table for the cosine (upper) and sine (lower) coefficients used to compute the frequency domain expansion coefficients $c_{nm}(t,t')$ and $s_{nm}(t,t')$. The horizontal axis shows the successive time samples, while the vertically offset lines show increments of the time derivative.}
\end{figure}

As a binary system approaches merger, the frequency evolution accelerates rapidly and the time domain approach becomes inefficient as it has to cover a larger and larger range of frequency derivatives. Rather than expanding the phase in $df/dt$, it becomes more efficient to switch to the frequency domain and expand the phase in $dt/df$:
 \begin{eqnarray}
 \Theta(f) &=&   \Theta(f_m) + 2 \pi (f-f_m)  t(f_m) \nonumber \\
 && + \pi (f-f_m)^2 t'(f_m)  + \dots \nonumber \\
 {\cal A}(f) &= &{\cal A}(f_m) + (f-f_m) {\cal A}'(f_m)  + \dots 
 \end{eqnarray}
where $f_m = m \Delta F$, and the expressions for $t(f)$ and $t'(f)$ are given in equations (\ref{spa1}) and (\ref{spa2}). To leading order in the amplitude and second order in the phase expansion, the wavelet expansion coefficients are given by
\begin{eqnarray}
&& w_{nm} = {\cal A}(f_m) \left( c_{nm}(t(f_m),t'(f_m)) \cos  \Theta(f_m) \right. \nonumber \\
&& \hspace*{0.8in} \left. -  s_{nm}(t(f_m),t'(f_m)) \sin\Theta(f_m) \right)
\end{eqnarray}
where
\begin{equation}
 c_{nm}(t,t') =  \int df \, \tilde{g}_{nm}(f) \cos(2 \pi (f-f_m)  t + \pi (f-f_m)^2 t' ) 
\end{equation}
and similarly for $s_{nm}(t,t')$ with cosine replaced by sine. Since the WDM wavelets are uniform in time and frequency, the coefficients need only be computed for a single reference frequency band and time slice.  The lookup table is generated on a grid spaced by $\delta t = T_w/M$ with $M \approx 400$, and $\delta  t'= \epsilon/(\Delta F^2)$ with $\epsilon \simeq 0.1$.   Figure~\ref{fig:taylorf} illustrates the frequency domain expansion coefficients $c_{nm}(t,t')$, $s_{nm}(t,t')$ as a function of the instantaneous time (across the horizontal) and time derivative (vertically offset lines) of the signal. 
The number of time derivatives in the table is chosen so as to cover a maximum time derivative $t'_{\rm max}$ that overlaps with the time domain expansion. Writing $t'_{\rm max} = a /\Delta F^2$, $\dot f_{\rm max} = b \Delta F/T_w$ and demanding that $\dot f_{\rm max}  \geq 1/t'_{\rm max}$ to ensure overlap of the two expansions, leads to the requirement $a b \geq \Delta F  \, T_w = q$.  The lookup tables used here have $q=16$, $a=8$ and $b=8$. The number of time samples required depends on the filter duration, $T_w$, and time evolution of the signal which increases with increasing $t'$. The total number of time samples scales as $(T_w + t'_{\rm max}\Delta F)/\delta t$. Since $(T_w + t' \Delta F) > \Delta T$, the wavelet coefficients for frequency band $m$ span at least $2q$ time slices.

\subsection{Waveform Test Case}\label{wtc}

To test the fast wavelet domain transforms it is helpful to work with a simple, analytically tractable waveform model. A good example is a ``chirplet'' - Gaussian enveloped sinusoid with linearly evolving frequency~\cite{mannVI91,Millhouse_2018}. In the frequency domain, a chirplet with central time $t_p$, central frequency $f_p$, time extent $\tau$ and frequency extent $\gamma$ have frequency domain phase and amplitude
\begin{eqnarray}\label{chirpf}
\Theta(f) &=& \Theta_0 + 2\pi f t_p + \pi \frac{\tau}{\gamma} (f-f_p)^2 \, , \nonumber \\
{\cal A}(f) &=& {\cal A}_0 \, e^{- (f-f_p)^2 /(2 \gamma^2)}\, .
\end{eqnarray}
The time-frequency mapping follows from equation (\ref{spa1}):
\begin{equation}
\frac{t-t_p}{\tau} = \frac{f-f_p}{\gamma}  \, .
\end{equation}
The time domain signal can be found by using the inverse stationary phase approximation:
\begin{equation}
h(t) = {\cal A}(f(t)) \sqrt{\frac{2\pi}{| \Theta^{''}(f(t))|}} \, e^{i (2\pi f(t) t + \Theta(f(t)) + \pi/4)}\, .
\end{equation}
The time domain amplitude and phase evaluate to
\begin{eqnarray}\label{chirpt}
 \Psi(t) &=&  2\pi f_p (t-t_p)  + \pi \frac{\gamma}{\tau} (t - t_p)^2+ \Theta_0 + \pi/4 \nonumber \\
A(t) & = & \sqrt{\frac{\gamma}{\tau}} {\cal A}_0 \, e^{-(t-t_p)^2/(2 \tau^2)}  \, .
\end{eqnarray}
The expression is exact up to fractional corrections of order $\vert {\cal A}''(f)/({\cal A}(f) \Theta''(f)) \vert \simeq 1/(4\pi \gamma \tau)$. The quantity $\gamma\tau$ is proportional to the time-frequency volume of the chirplet, indicating that the stationary phase approximation improves as the time-frequency volume increases.

\begin{figure}[htp]
\includegraphics[width=0.48\textwidth]{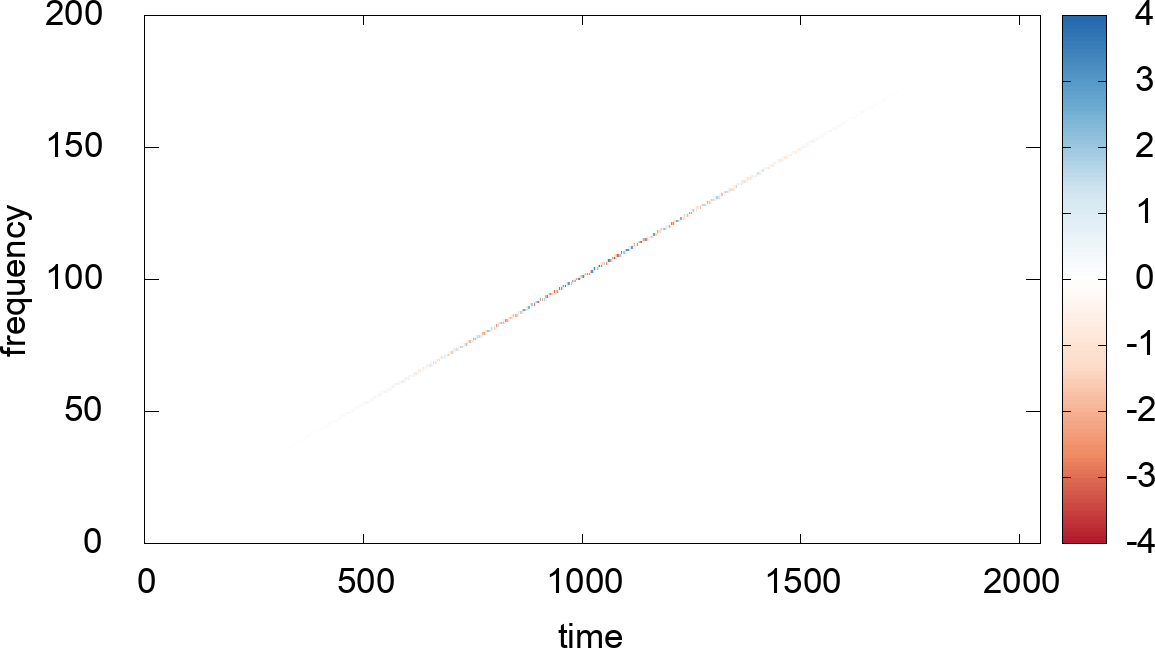} 
\includegraphics[width=0.48\textwidth]{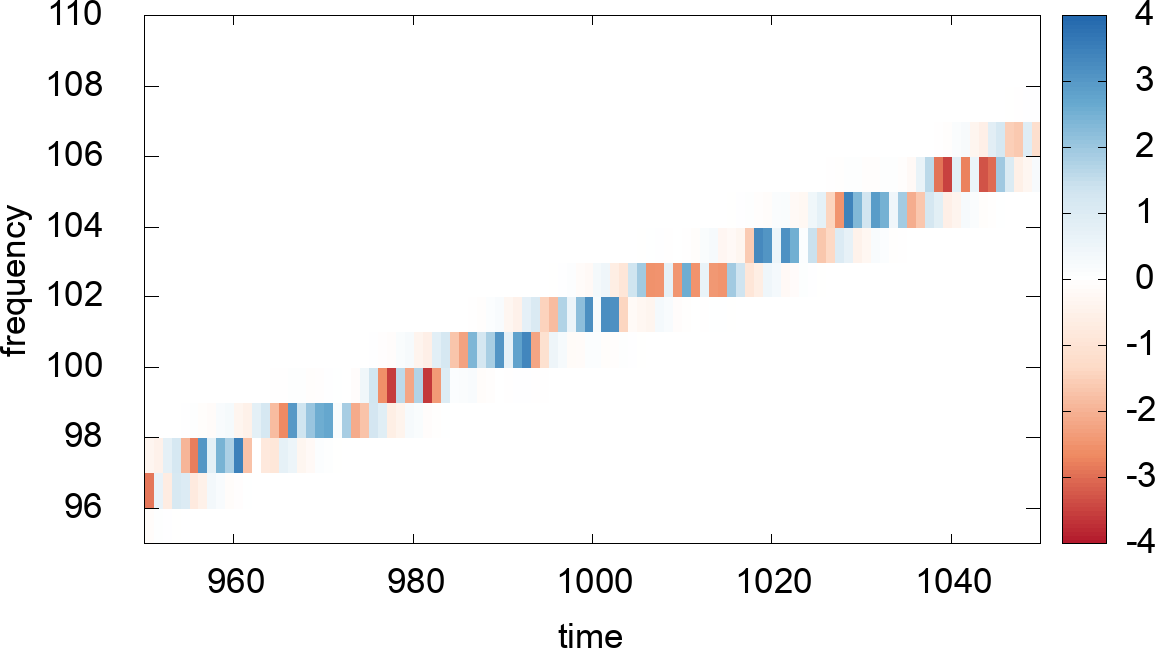} 
\caption{\label{fig:test} The WDM transform of a chirplet. Time is in units of $\Delta T$ and frequency is in units of $\Delta F$. The upper panel shows the full signal, while the lower panel zooms in on a smaller segment. The color bar indicates the wavelet amplitudes.}
\end{figure}

Figure~\ref{fig:test} shows the WDM transform of a chirplet with parameter $t_p = 1024\, \Delta T$, $f_p = 103.4\, \Delta F$, ${\cal A}_0 = 10000$, $ \Theta_0  = 0.0$,  $\dot f = \gamma/\tau = 3.1 \Delta F/T_w$ and $\gamma = f_p/4$. The wavelet transform has $N_t=2048$ time slices and $N_f=2048$ frequency bands. Only the frequency bands occupied by the signal are shown in Figure~\ref{fig:test}. The un-weighted match between two wavelet transform $w_{nm}$, $u_{nm}$ can be defined as
\begin{equation}
 {\rm M} = \frac{ ( w| u) }{\sqrt{(w|w)(u|u)}}
\end{equation}
where
\begin{equation}
 ( w| u) = \sum_{n,m} w_{nm} u_{nm} \ .
 \end{equation}
The mis-match ${\rm MM} = 1 - {\rm M}$ between the direct WDM transform and the various fast approximations was ${\rm MM}=6 \times 10^{-10}$ for the sparse time domain transform, ${\rm MM}=4 \times 10^{-6}$ for the sparse frequency domain transform, ${\rm MM}=3 \times 10^{-6}$ for the Taylor expanded time domain transform, and ${\rm MM}=3 \times 10^{-5}$ for the Taylor expanded frequency domain transform. In this instance the fast frequency domain transforms are slightly less accurate than the fast time domain transforms since the data was generated using the time domain expression (\ref{chirpt}), while the frequency domain transforms use the frequency domain expression (\ref{chirpf}). These differ slightly since only the lowest order stationary phase approximation was used to map between the time and frequency domain expressions. If necessary, the accuracy of the Taylor expanded WDM transforms can be improved by including the leading order amplitude evolution.
Using a single 2.9 GHz CPU core, the direct time domain transform took 5.6 s, and the direct frequency domain transform took 0.9 s. These numbers do not include the 4.1 s it took to generate the full time domain waveform. The sparse time domain transform took 15 ms, the sparse frequency domain transform took 2 ms, the Taylor expanded time domain transform took 0.7 ms and the Taylor expanded frequency domain transform took 0.9 ms. 

 \subsection{Fast Inspiral-Merger-Ringdown Transform}\label{IMR}
 
\begin{figure}[htp]
\includegraphics[width=0.5\textwidth]{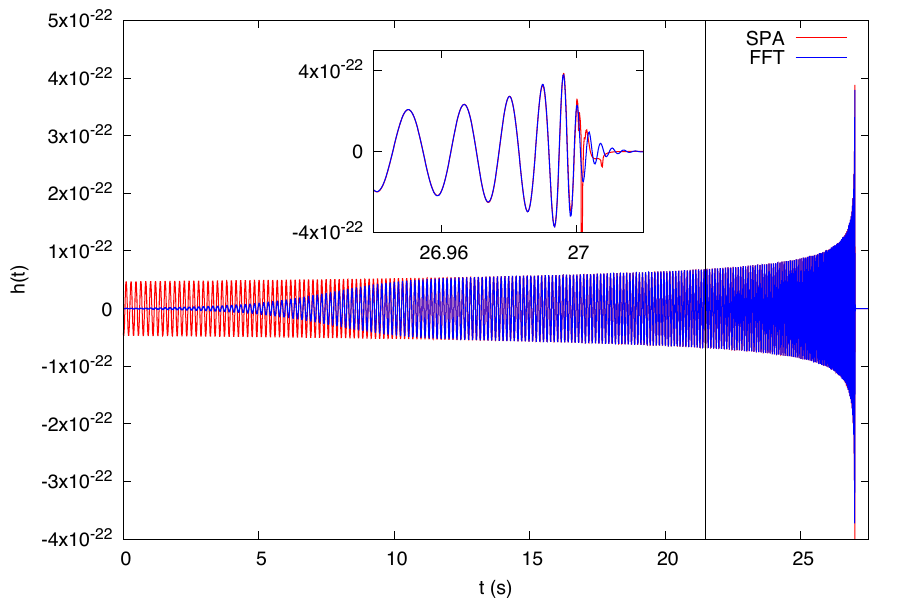} 
\caption{\label{fig:BHwave} Construction of the reference time-domain signal for a binary black hole merger. At early times the signal is computed using the stationary phase approximation, while at late times the inverse Fourier transform is used. The transition between the two regimes is marked by a vertical black line. The inset shows the breakdown of the SPA at merger.}
\end{figure}

 The time domain and frequency domain fast wavelet transform techniques can be combined together to yield a fast wavelet transform for inspiral-merger-ringdown models. As an illustrative example, consider the frequency domain ``PhenomD'' phenomenological model~\cite{Khan:2015jqa}, which describes the dominant $\ell = |m| = 2$ mode of a quasi-circular, spin-aligned binary system using a combination of the TaylorF2 post-Newtonian model for the inspiral and a black hole perturbation theory model for the ringdown, bridged by a fit to a suite of numerical relativity waveforms describing the final inspiral and merger.  The PhenomD model returns expression for the frequency domain amplitude and phase,  ${\cal A}(f), \Theta(f)$. A numerical implementation of the stationary phase approximation (SPA) can be used to find the time domain amplitude and phase $A(t), \Psi(t)$. The SPA breaks down at merger, but this is not a limitation since the time domain expressions are only used at early times where the signal is slowly evolving. To generate a reference time domain signal to be used with the direct wavelet transform the SPA waveform is spliced together with the inverse fast Fourier transform (FFT) of the frequency domain signal, which has to be smoothly tapered to zero at low frequencies to avoid Gibbs oscillations in the time domain. The splicing procedure is illustrated in Figure~\ref{fig:BHwave}. Note that the SPA and the FFT match perfectly in amplitude and phase at the transition point.
 
\begin{figure}[htp]
\includegraphics[width=0.5\textwidth]{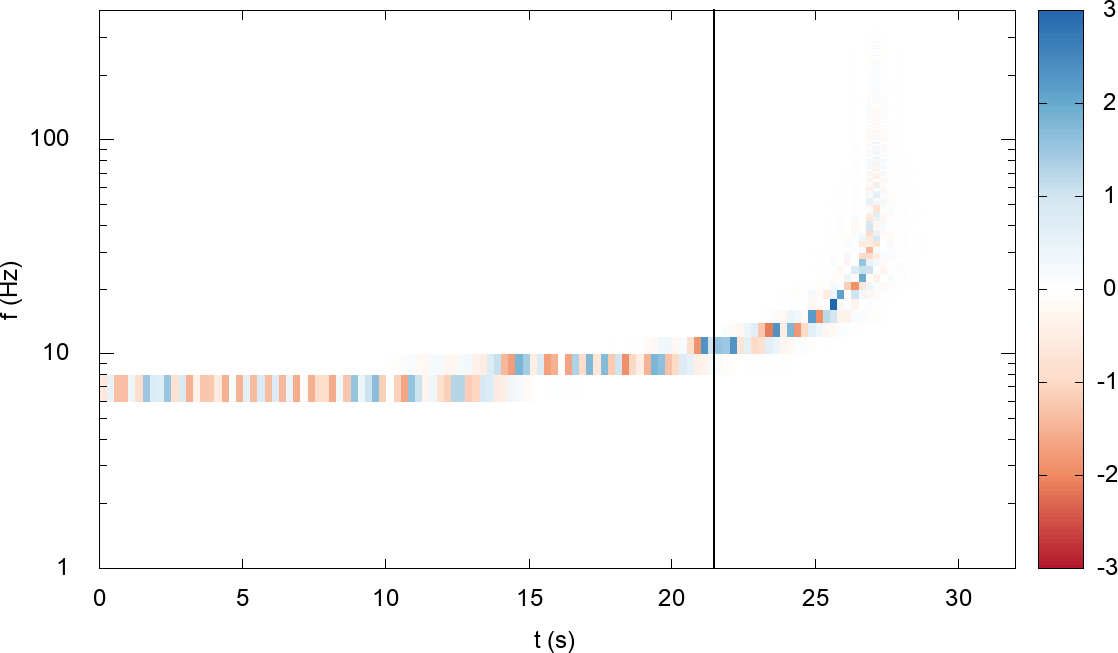} 
\caption{\label{fig:BHwdm} The WDM wavelet transform of the binary black hole signal shown in Figure~\ref{fig:BHwave}. The vertical black line marks the transition from the time domain Taylor transform to the sparse frequency domain transform.}
\end{figure}

A fast wavelet transform of the PhenomD waveform can be constructed by combining the time domain Taylor expansion for early times and the sparse frequency domain transform for late times. The transition time is chosen such that the times $t_m$, $t_{m+1}$ corresponding to successive frequency bands $m$, $m+1$ are separated by no more than half-width of the sparse transform: $t_{m+1} - t_m < q\Delta T$. This condition ensures that the sparse frequency domain transform covers the time slices occupied by the signal. The binary black hole systems shown in Figure~\ref{fig:BHwave} had masses $m_1=35 M_\odot$, $m_2=30 M_\odot$, dimensionless spins $\chi_1=0.3$, $\chi_2=0.1$ and merger time $t_c=27$ s. For this system the transition from the time domain transform to the frequency domain transform occurred at $t_* = 21.5$ s. The transform was computed using $N_f = 1024$, $N_t=128$, $\Delta t = 1/4096$ and covered 32 s. The wavelet transform of the signal is shown in Figure~\ref{fig:BHwdm}. The mis-match between the direct wavelet transform of the signal and the fast wavelet transform was ${\rm MM} = 1.05 \times 10^{-4}$. The match could be improved by including time evolution of the amplitude in the time domain Taylor expansion. The fast wavelet transform took 1.3 ms. This is a significant saving compared to the native frequency domain waveform, which took 8.5 ms to generate. The wavelet transform covered $N_p = 9339 \approx 26 \sqrt{N}$ pixels.
 
\begin{figure}[htp]
\includegraphics[width=0.5\textwidth]{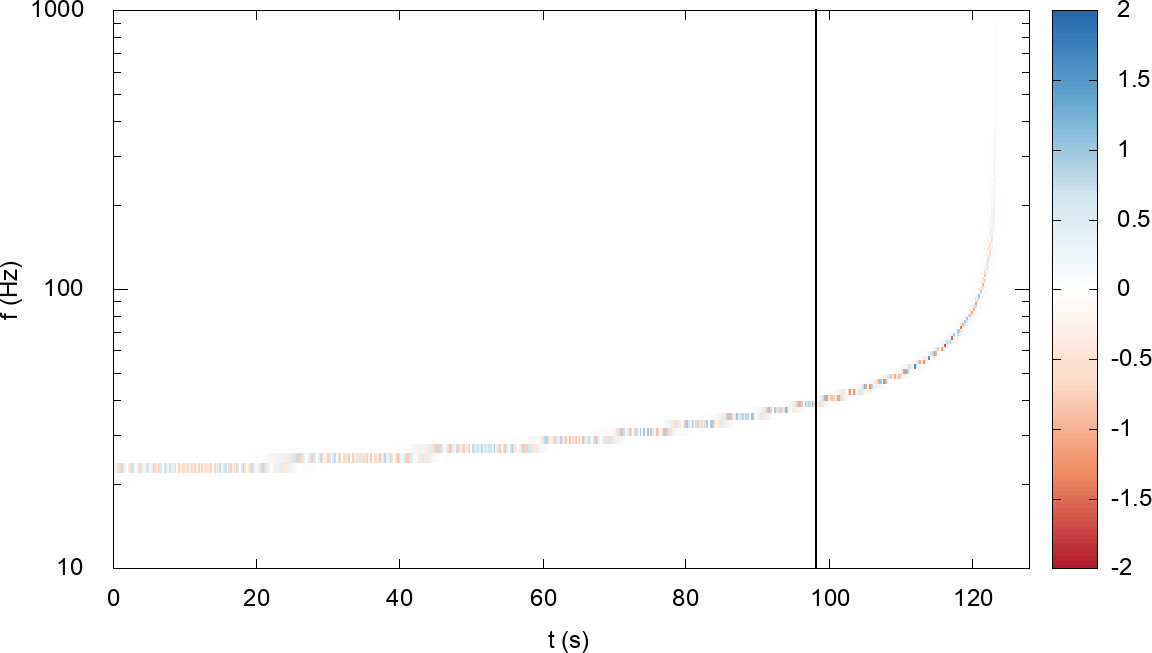} 
\caption{\label{fig:BNSwdm} The WDM wavelet transform of a binary system with neutron star masses and spins. The vertical black line marks the transition from the time domain Taylor transform to the sparse frequency domain transform.}
\end{figure}

 The computational savings increase for lower mass systems. For example, consider a system with neutron star masses and spins: $m_1=1.6 M_\odot$, $m_2=1.4 M_\odot$, $\chi_1=0.02$, $\chi_2=0.03$ and merger time $t_c=123$ s. The WDM transform of this signal is shown in Figure~\ref{fig:BNSwdm} for $N_f = 1024$, $N_t=512$, $\Delta t = 1/4096$ covering 128 s. In this case the transition between the time domain Taylor expansion and the sparse frequency domain transform occurred at  $t_* = 98.25$ s. The mis-match between the direct wavelet transform of the signal and the fast wavelet transform was ${\rm MM} = 3.6 \times 10^{-6}$. The
 fast wavelet transform took 4.0 ms, which is significantly less than the 45 ms for the native frequency domain signal.  The wavelet transform covered $N_p = 32843 \approx 45 \sqrt{N}$ pixels.
 
 \subsection{Fast LISA Ultra Compact Binary Transforms}
 
Ultra compact galactic binaries (UCBs) are expected to be the most common source detected by future space based gravitational wave detectors operating in the mHz regime~\cite{1990ApJ...360...75H}. Most UCBs are expected to be on slowly evolving, quasi-circular orbits, producing signals that can be accurately modeled using leading order post-Newtonian waveforms. For detectors such as LISA, which orbit the Sun and have time-dependent response functions, the UCB signals pick up amplitude and frequency modulations that vary on the orbital timescale. Wavelet transforms in the mHz regime will typically have pixels that are several hours in extent and hundredths of a mHz in bandwidth. The modulated UCB signals are effectively constant across the wavelet pixels, and can be accurately modeled using the leading order fast time domain Taylor transform. 
The $A,E,T$ time delay interferometry channels can each be expressed in the form $h_I(t) = A_I(t) \cos \Psi_I(t)$. In the low frequency limit the expressions have the form~\cite{Cutler:1997ta}
\begin{eqnarray}
A_I(t) &=& \left[ A_+^2(t) F_{I+}^2(t) + A_\times^2(t) F_{I\times}^2(t)\right]^{1/2} \nonumber \\
\Psi_I(t) &=& \int 2\pi f(t) dt + \Psi_{Ip}(t) + \Psi_{Id}(t) \nonumber \\
f(t) &=& f_0 + \dot f_0 t + \frac{1}{2} \ddot f_0 t^2 \, ,
\end{eqnarray}
with amplitudes
\begin{eqnarray}
A_+(t) &=& \frac{4 {\cal M}^{5/3} (\pi f(t))^{2/3} }{D_L}\left(\frac{1+\cos^2\iota}{2}\right)  \nonumber \\
A_\times(t) & =&   \frac{4 {\cal M}^{5/3} (\pi f(t))^{2/3} }{D_L} \cos\iota
\end{eqnarray}
and polarization and Doppler phases
\begin{eqnarray}
\Psi_{Ip}(t) &=& {\rm atan} \left(\frac{ -A_{I\times}(t)F_{I\times}}{ A_+(t) F_{I+}(t)} \right)\nonumber \\
\Psi_{Id}(t) &=& 2 \pi f(t) {\rm AU} \sin\theta \cos(2\pi t/{\rm yr} + \phi_0 - \phi)\, .
\end{eqnarray}
Here ${\cal M}$ is the chirp mass, $D_L$ is the luminosity distance, $\iota$ is the orbital inclination relative to the line off sight, and $\theta,\phi$ are the sky location of the source in ecliptic coordinates. The time dependent antenna patterns $F_{I+}(t), F_{I\times}(t)$ depend on the sky location and polarization angle of the source. Similar, yet more complicated expressions for the time dependent amplitude and phase can be found for the full response function, taking into account finite armlength effects - see appendix B of Ref.~\cite{Cornish:2020vtw} for details. The results shown here utilize the full LISA instrument response.

To leading order, the intrinsic frequency change for a UCB scales as
\begin{equation}
\dot f = 3.255\times 10^{-16} \left(\frac{f}{3\, {\rm mHz}}\right)^{11/3} \left(\frac{{\cal M}}{1 M_\odot}\right)^{5/3} \, {\rm Hz} \,{\rm s}^{-1}
\end{equation}
while the magnitude of the maximum Doppler shift scales as
\begin{equation}
\dot f_d = 5.93\times 10^{-14} \left(\frac{f}{3\, {\rm mHz}}\right) \sin\theta  \, {\rm Hz} \,{\rm s}^{-1} \, .
\end{equation}
For typical choices of $\Delta T \sim 10^5 \rightarrow 10^6 \, s$ and $\Delta F \sim 1 \rightarrow 10\, \mu{\rm Hz}$, the fractional frequency change across the window function, $\dot f T_w/\Delta F$, is of order $10^{-2}$ or less. Consequently, the lookup table for UCBs only has to cover zeroth order in frequency change across the filter.

 \begin{figure}[htp]
\includegraphics[width=0.5\textwidth]{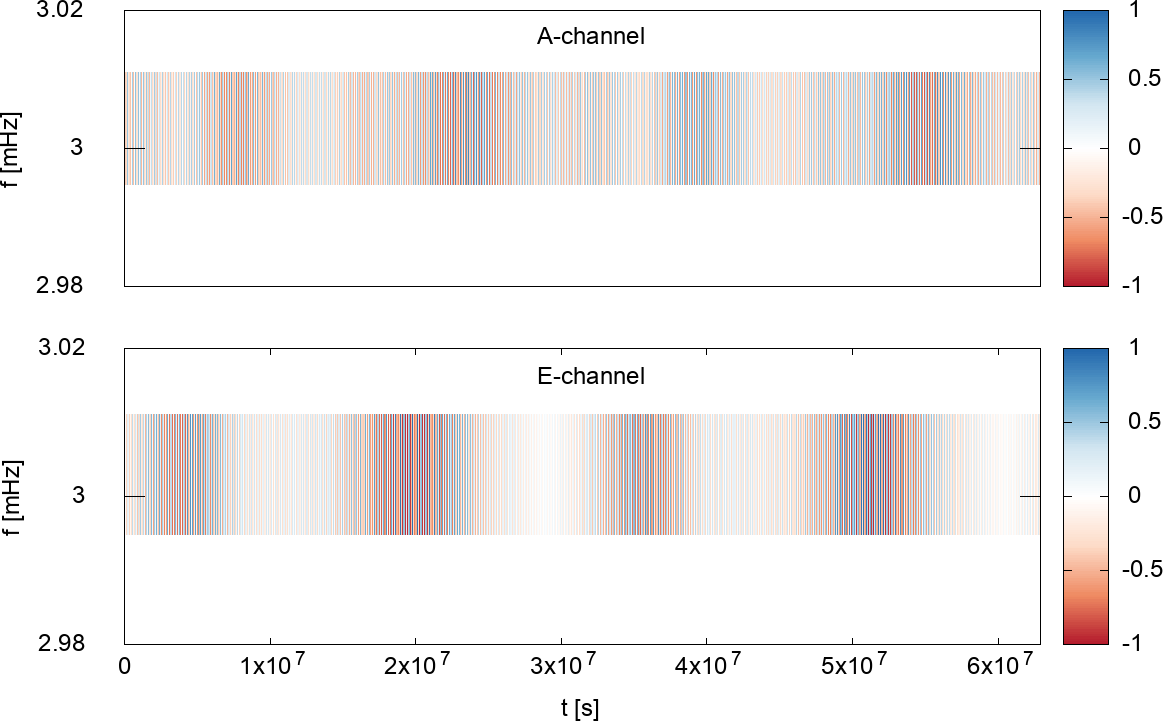} 
\caption{\label{fig:UCBwdm} The WDM wavelet transform of the LISA response to a galactic binary in the TDI A and E channels. The color scale is in units of $10^{-20}$.}
\end{figure}

Figure~\ref{fig:UCBwdm} shows the WDM transform of the LISA response to a galactic binary with $f_0 = 3$ mHz, $\dot f_0 = 2.584\times 10^{-16} \, {\rm Hz} {\rm s}^{-1}$, sky location $\cos\theta = 0.2$, $\phi= 1.0$, orientation $\cos \iota = -0.3$, $\psi = 0.8$, initial phase $\Psi_0=1.2$ and amplitude $A_0 = 1.96 \times 10^{-21}$.
These parameters are encoded into the ``bar-code'' structure of the WDM transform. Each galactic binary will have a unique bar-code, and even though many thousands of binaries will occupy a given frequency band of the transform, their unique encodings allow the individual signals to be extracted, much like a television cable box decodes the multiplexed signals carried on a single co-axial cable. The transform used $N_f=N_t=2048$, and the data had a sample cadence of $\Delta t = 15$ s and a total duration of roughly two years. The wavelet pixels have duration $\Delta T = 30720$ s and bandwidth $\Delta F = 1.627604\times 10^{-2}$ mHz. Generating both channels took a total of 1.3 ms, which is slightly faster than the fast frequency domain transform~\cite{Cornish:2007if} that is currently used in LISA data analysis studies~\cite{Littenberg:2020bxy}. The mis-match between the fast and full transforms was ${\rm MM} = 4.6 \times 10^{-7}$ for the A-channel and ${\rm MM} = 1.1 \times 10^{-6}$ for the E-channel.

\begin{figure}[htp]
\includegraphics[width=0.5\textwidth]{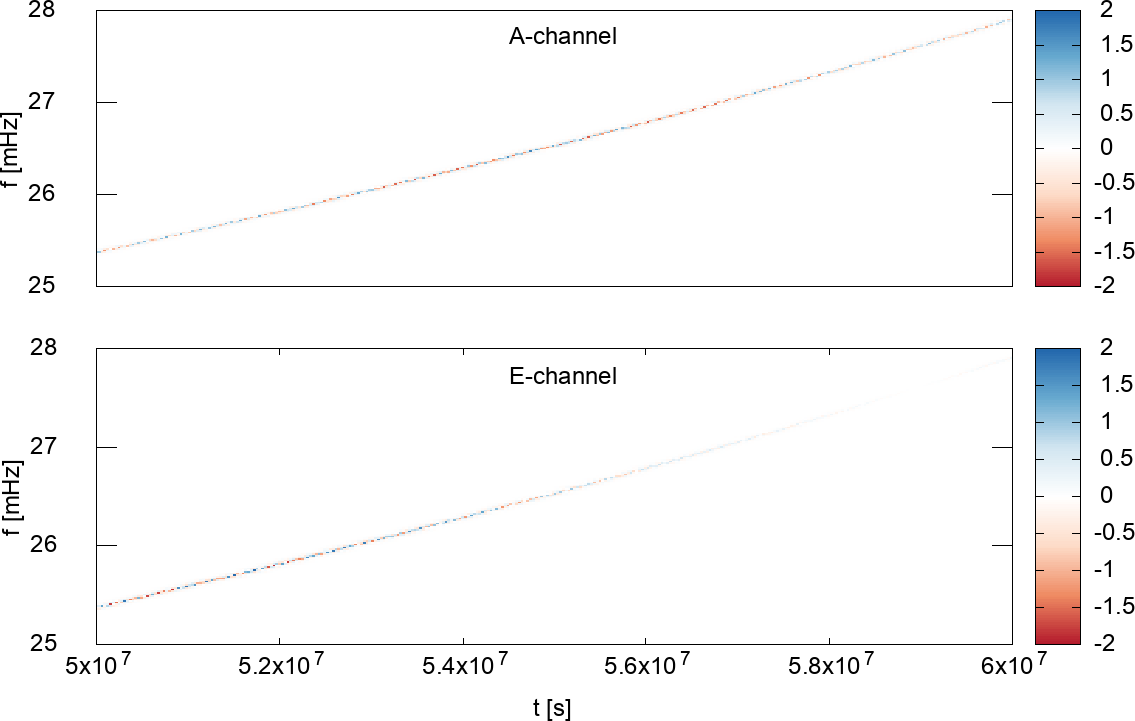} 
\caption{\label{fig:SOBHwdm} The WDM wavelet transform of the LISA response to a stellar origin black hole binary in the TDI A and E channels. The color scale is in units of $10^{-22}$.}
\end{figure}

The UCB transform can be extended to include stellar origin black holes (SOBHs) by using a high order post-Newtonian inspiral model, and by including frequency derivatives in the lookup table. Using the PhenomD waveform model, and considering the same SOBH system as before, but starting 3 years before merger, yields the LISA signal shown in Figure~\ref{fig:SOBHwdm}. To make the signal track more visible, only the final 4 months of the full two years of simulated LISA data are shown. The fast WDM waveform generation took 4.6 ms. This is a significant saving compared to the 6.8 seconds it takes to compute the waveform in the frequency domain. The mis-match between the fast and full transforms was ${\rm MM} = 8.1 \times 10^{-6}$ for the A-channel and ${\rm MM} = 1.6 \times 10^{-5}$ for the E-channel. Here only the time domain Taylor expansion was used. To achieve higher accuracy, and to cover systems with larger chirp masses and/or shorter times until merger, it is necessary to switch to the sparse frequency domain response once the frequency evolution is much larger than $\Delta F/T_w$,  as was done for the BH mergers described in \S \ref{IMR}.

\section{Summary}

A time-frequency approach to gravitational wave data analysis using discrete wavelet transforms offers many advantages over the traditional time or frequency domain approaches. Wavelet based analyses are well suited for modeling non-stationary instrument noise.  Wavelet domain models of binary merger signals are often significantly faster to compute than time or frequency domain models. Together these advantages make a strong case for moving gravitational wave data analysis to the wavelet domain.

To facilitate the transition, a simple translator is being developed for converting existing waveforms to the wavelet domain.  Ready to use algorithms for estimating the evolutionary power spectra are also being developed. The translator takes as input the amplitude and phase of each harmonic, either in the time domain or the frequency domain, then based on the frequency evolution, determines which fast transform to apply. The transform returns the non-zero wavelet coefficients as a sparse array with an indexing table to facilitate fast likelihood calculations. A range of tools for estimating the evolutionary power spectrum are also being developed, including simple smoothed median estimators like the ones described here, in addition to more sophisticated adaptive (trans-dimensional) Bayesian methods.

\section*{Acknowledgments}
This work was supported by the NASA LISA foundation Science Grant 80NSSC19K0320 and NSF awards PHY1607343 and PHY1912053. This work was initiated while the author was on sabbatical at the Observatoire de la C\^{o}te d'Azur, kindly hosted by Nelson Christensen, and supported in part by the Centre National d'\'{E}tudes Spatiales. The author benefited from discussions with Will Farr, Sergey Klimenko and Guy Nason. This research has made use of data obtained from the Gravitational Wave Open Science Center (https://www.gw-openscience.org), a service of LIGO Laboratory, the LIGO Scientific Collaboration and the Virgo Collaboration. LIGO is funded by the U.S. National Science Foundation. Virgo is funded by the French Centre National de Recherche Scientifique (CNRS), the Italian Istituto Nazionale della Fisica Nucleare (INFN) and the Dutch Nikhef, with contributions by Polish and Hungarian institutes.

\bibliography{refs}

\end{document}